\documentclass[reprint,superscriptaddress,preprintnumbers,longbibliography,
amsmath,amssymb,aps,floatfix,prb,twocolumn, tightenlines 
]{revtex4-2}
\usepackage[colorlinks=true,
citecolor=blue,
linkcolor=black,
anchorcolor=black,
urlcolor=magenta]{hyperref}
\usepackage{amsmath}
\usepackage{amsfonts}
\usepackage{amssymb}
\usepackage{braket}
\usepackage{graphicx}
\usepackage{siunitx}
\usepackage{svg}

\newcommand{\subheading}[1]{\par\medskip\noindent\textbf{#1}\par\smallskip}

\newcommand{\addrA}{Beijing National Laboratory for Condensed Matter Physics,\\ Institute of Physics, Chinese Academy of Sciences, Beijing 100190, China} 
\newcommand{\addrB}{School of Physical Sciences, University of Chinese Academy of Sciences, Beijing 100049, China} 
\newcommand{\addrC}{Beijing Key Laboratory of Fault-Tolerant Quantum Computing,\\ Beijing Academy of Quantum Information Sciences, Beijing 100193, China} 
\newcommand{\addrD}{Hefei National Laboratory, Hefei 230088, China} 
\newcommand{\addrE}{Beijing Key Labratory of Advanced Quantum Technology, Beijing 100190, China} 
\newcommand{\addrF}{Fujian Key Laboratory of Quantum Information and Quantum Optics, College of Physics and \\
	Information Engineering, Fuzhou University, Fuzhou, Fujian 350108, China}

\begin{document}
	\title{Programmable spectral symmetries in an anisotropic quantum Rabi simulator}
	
	\author{Jia-Cheng Song}
	\thanks{These authors contributed equally to this work.}
	\affiliation{\addrA}
	\affiliation{\addrB}
	
	\author{Yu Liu}
	\email{yuliu@iphy.ac.cn}
	\thanks{These authors contributed equally to this work.}
	\affiliation{\addrA}
	
	\author{Ming-Chuan Wang}
	\thanks{These authors contributed equally to this work.}
	\affiliation{\addrA}
	\affiliation{\addrB}
	
	\author{Ke-Xiong Yan}
	\thanks{These authors contributed equally to this work.}
	\affiliation{\addrF}
	
	\author{Yang He}
	\affiliation{\addrA}
	\affiliation{\addrB}
	
	\author{Yun-Hao Shi}
	\affiliation{School of Physics, Xi’an Jiaotong University, Xi’an 710049, China}
	
	\author{Wei-Ping Yuan}
	\affiliation{\addrA}
	\affiliation{\addrB}
	
	\author{Cheng-Lin Deng}
	\affiliation{\addrC}
	
	\author{Li Li}
	\affiliation{\addrA}
	\affiliation{\addrB}
	
	\author{Zhen-Ting Bao}
	\affiliation{\addrA}
	\affiliation{\addrB}
	
	\author{Yutao Chen}
	\affiliation{\addrA}
	\affiliation{\addrB}
	
	\author{Xu-Yang Gu}
	\affiliation{\addrA}
	\affiliation{\addrB}
	
	\author{Tian-Ming Li}
	\affiliation{\addrA}
	\affiliation{\addrB}
	
	\author{Gui-Han Liang}
	\affiliation{\addrA}
	
	\author{Zheng-He Liu}
	\affiliation{\addrA}
	\affiliation{\addrB}
	
	\author{Wei-Guo Ma}
	\affiliation{\addrA}
	\affiliation{\addrB}
	
	\author{Zhen-Yu Peng}
	\affiliation{\addrA}
	\affiliation{\addrB}
	
	\author{Shuai-Li Wang}
	\affiliation{\addrA}
	\affiliation{\addrB}
	
	\author{Yong-Xi Xiao}
	\affiliation{\addrA}
	\affiliation{\addrB}
	
	\author{Yi-Han Yu}
	\affiliation{\addrA}
	\affiliation{\addrB}
	
	\author{Jia-Chi Zhang}
	\affiliation{\addrA}
	\affiliation{\addrB}
	
	\author{Kui Zhao}
	\affiliation{\addrC}
	
	\author{Min-Xuan Zhou}
	\affiliation{\addrA}
	\affiliation{\addrB}
	
	\author{Kaixuan Huang}
	\affiliation{\addrC}
	
	\author{Yu-Ran Zhang}
	\affiliation{School of Physics and Optoelectronics, South China University of Technology, Guangzhou 510640, China}
	
	\author{Yu-Xiang Zhang}
	\affiliation{\addrA}
	\affiliation{\addrB}
	\affiliation{\addrD}
	
	\author{Zhongcheng Xiang}
	\affiliation{\addrA}
	\affiliation{\addrB}
	\affiliation{\addrD}
	\affiliation{\addrE}
	
	\author{Dongning Zheng}
	\affiliation{\addrA}
	\affiliation{\addrB}
	\affiliation{\addrD}
	\affiliation{\addrE}
	
	\author{Ye-Hong Chen}
	\email{yehong.chen@fzu.edu.cn}
	\affiliation{\addrF}
	\affiliation{Quantum Information Physics Theory Research Team, Center for Quantum Computing, RIKEN, Wako-shi, Saitama 351-0198, Japan}
	
	\author{Kai Xu}
	\email{kaixu@iphy.ac.cn}
	\affiliation{\addrA}
	\affiliation{\addrB}
	\affiliation{\addrC}
	\affiliation{\addrD}
	\affiliation{\addrE}
	
	\author{Heng Fan}
	\email{hfan@iphy.ac.cn}
	\affiliation{\addrA}
	\affiliation{\addrB}
	\affiliation{\addrC}
	\affiliation{\addrD}
	\affiliation{\addrE}

	\date{\today}

	\begin{abstract}
		The quantum Rabi model captures fundamental aspects of light--matter interaction, where symmetry dictates both spectra and dynamics. Over the past years, experiments have explored many of its nonperturbative properties, but have mostly focused on the isotropic limit, where rotating and counterrotating processes are locked together, leaving the broader symmetry landscape largely unexplored.
		Here we realize a programmable anisotropic quantum Rabi model in a superconducting processor, with independent control of the rotating and counterrotating couplings $(g_1,g_2)$ and of a transverse bias $\varepsilon$. Continuous anisotropy tuning, combined with a duality mapping, gives access to the full parameter space from the Jaynes-Cummings to the anti-Jaynes-Cummings limits. In the deep-strong-coupling regime, we show that anisotropy reconstructs the spectrum and turns complete collapse-revival dynamics into incomplete revivals even near degeneracy. With adiabatic state preparation and joint tomography, we resolve an anisotropy-induced ground-state parity switch, a crossing that has no analogue in the isotropic model. We further observe selective tunnelling associated with hidden symmetry in biased Rabi models and track its anisotropic displacement within the same device. These results establish a controllable route to engineering nonperturbative light--matter Hamiltonians, where symmetry, spectrum, and dynamics can be programmed independently.
	\end{abstract}
	\maketitle
	
	The quantum Rabi model (QRM)~\cite{Rabi1936}, a two-level system coupled to a single bosonic mode, is one of the simplest Hamiltonians, in which the light--matter interaction becomes genuinely quantum. It underlies phenomena ranging from coherent vacuum Rabi oscillations~\cite{Brune1996,Wallraff2004,Blais2021} to the generation of nonclassical states~\cite{Hofheinz2009,Leghtas2015}. When the interaction strength $g$ is small compared with the mode frequency $\omega$, the Jaynes-Cummings model (JCM)~\cite{Jaynes1963} provides an accurate description, because the total excitation number is conserved. This conservation law produces the characteristic $\sqrt{n}$ ladder~\cite{Brune1996,Deppe2008} and coherent matter-field exchange~\cite{Hofheinz2009, Krantz2019, Lled2023}. As the coupling approaches the mode frequency, the rotating-wave approximation (RWA) breaks down. The system then enters a nonperturbative regime~\cite{Wallraff2004, Niemczyk2010, FornDiaz2017, Yoshihara2017, Forn2019, FriskKockum2019, Qin2024}, where excitation manifolds hybridize, vacuum dressing becomes pronounced~\cite{FornDaz2010, Mornhinweg2024, Yoshihara2017}, and intrinsically multifrequency dynamics emerges~\cite{Eberly1980,Casanova2010}. These effects define the ultrastrong-coupling (USC) regime, typically when $g/\omega>0.1$, and the deep-strong-coupling (DSC) regime, often identified by $g/\omega\gtrsim 1$.
	Yet nonperturbative coupling strength is only one axis of the problem. The relative weight of rotating and counterrotating processes also determines the conserved quantities, selection rules, and level crossings of the model. Most experiments have nevertheless focused on the isotropic QRM~\cite{Braumuller2017, Langford2017, Lv2018, Chen2021, Cai2021, Wang2023, Zheng2023}, where this balance is fixed. As a result, central questions remain open: how does the anisotropy reorganize the spectrum, how do symmetry-protected crossings appear or disappear, and can hidden symmetries be detected dynamically within a single programmable device? Addressing these questions requires independent control of the two interaction channels, opening the access to a broader landscape beyond the RWA~\cite{Baksic2014,Grimsmo2013,Xie2014,Shen2017,Ye2025}.
	
	\begin{figure*}[!t]
		\centering
		\includegraphics[width=0.97\linewidth]{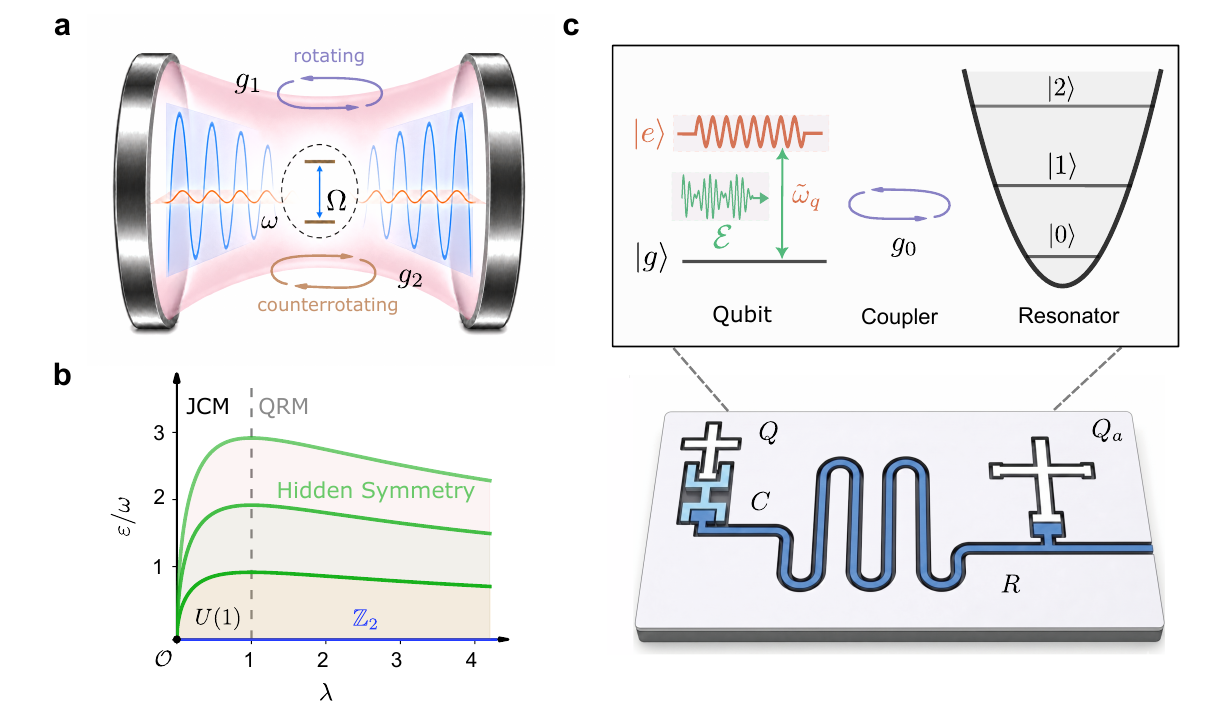}
		\caption{\textbf{Device and theoretical model}.
			\textbf{a,} Schematic illustration of the anisotropic quantum Rabi model, describing a qubit coupled to two quadratures of the quantized light field.
			\textbf{b,} Symmetry structure of the AQRM in the parameter space $(\lambda,\varepsilon/\omega)$. The point $(0,0)$ corresponds to the $U(1)$-symmetric JCM limit, the line $\varepsilon=0$ preserves the $\mathbb{Z}_2$ parity symmetry. Even outside these symmetry manifolds, hidden symmetries can emerge at special parameter points.
			\textbf{c,} Schematic diagram of the experimental device and pulse sequence. The system qubit $Q$ is coupled to the resonator while subject to longitudinal sinusoidal modulation and a multi-tone transverse drive. Together, these modulations realize an effective AQRM Hamiltonian.}
		\label{fig1}
	\end{figure*}
	
	A natural framework for this control is the anisotropic quantum Rabi model (AQRM)~\cite{Xie2014}, which extends the QRM by coupling a two-level system to both quadratures of a bosonic mode, as illustrated in Fig.~\ref{fig1}\textbf{a}. The AQRM provides a tunable setting for quantum criticality, quantum sensing, and nonclassical-state generation~\cite{Liu2017,Shen2017,Xie2020,Zhu2023,Lyu2024,Xu2024}.  The Hamiltonian takes the form (by setting $\hbar=1$)
	\begin{equation}\label{eqHamAQRM}
		\hat{H}=\frac{\Omega}{2}\hat{\sigma}_z+\omega \hat{a}^{\dagger}\hat{a}+g_1\!\left(\hat{H}_{\mathrm{r}}+\lambda \hat{H}_{\mathrm{cr}}\right)+\frac{\varepsilon}{2}\hat{\sigma}_x,
	\end{equation}
	where $\hat{H}_{\mathrm{r}}=\hat{a}^{\dagger}\hat{\sigma}_-+\hat{a}\hat{\sigma}_+$ and $\hat{H}_{\mathrm{cr}}=\hat{a}^{\dagger}\hat{\sigma}_+ + \hat{a}\hat{\sigma}_-$ are the rotating and counterrotating interactions, respectively.
	Here, $\Omega$, $\omega$, and $\varepsilon$ are the qubit frequency, mode frequency, and transverse bias, respectively. And $\lambda=g_2/g_1$ sets the relative strength of the two channels. Tuning $\lambda$ interpolates between the JCM ($\lambda=0$) and the isotropic QRM ($\lambda=1$). The AQRM exhibits a layered symmetry structure. For $\lambda=0$ and $\varepsilon=0$, it has a continuous $U(1)$ symmetry associated with the conservation of the total excitation number $\hat N=\hat a^\dagger \hat a+(\hat{\sigma}_z+1)/2$. For $\lambda\neq0$ but $\varepsilon=0$, this symmetry is reduced to a discrete $\mathbb{Z}_2$ parity symmetry generated by $\hat{\mathcal{P}}=-(-1)^{\hat a^\dagger \hat a}\hat{\sigma}_z$~\cite{Braak2011}. A finite bias generally breaks parity and turns protected crossings into avoided crossings, yet exact crossings can reemerge at the discrete commensurate biases associated with exceptional solutions and hidden symmetries~\cite{Judd1979,Tomka2014,Ashhab2020,Li2021, Mangazeev2021, ReyesBustos2021, Nguyen2024,Yang2025}. This symmetry structure is summarized in Fig.~\ref{fig1}b. In addition to this rich structure and broad interest, direct experimental access to the AQRM and systematic probes of its spectra and dynamics have remained challenging, despite some theoretical discussions~\cite{Skogvoll2021,Chen2024, Tabatabaei2025}.
	
		\begin{figure*}[!t]
		\centering
		\includegraphics[width=0.99\linewidth]{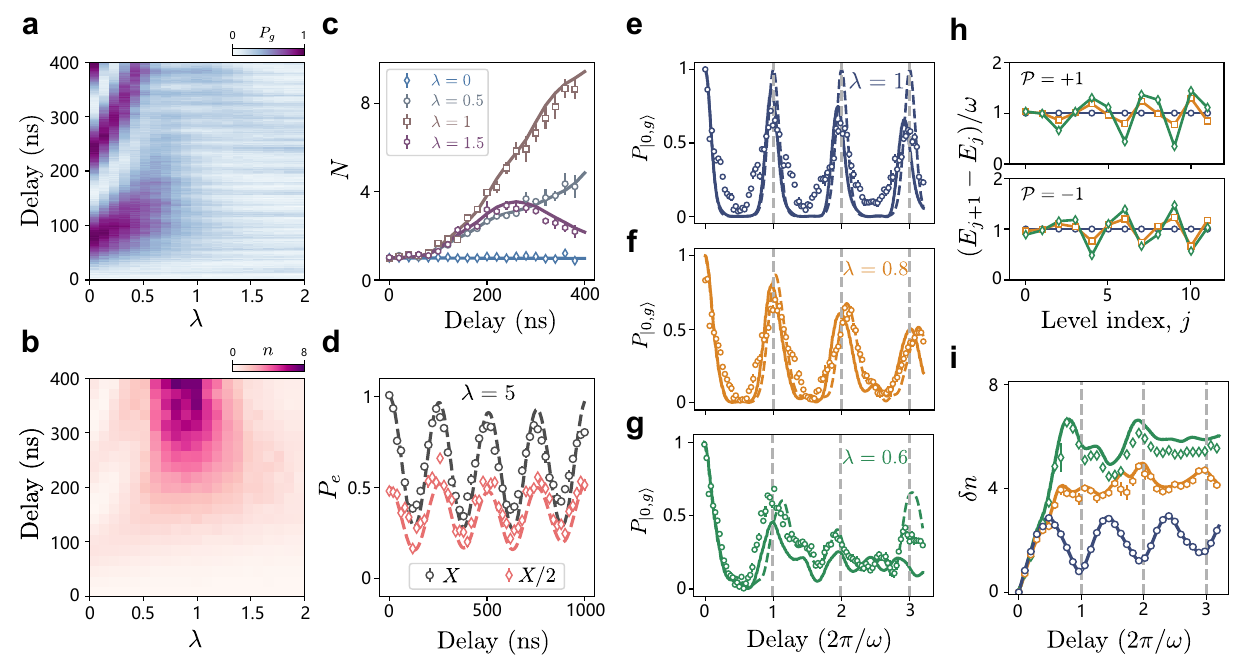}
		\caption{
			\textbf{Tunable anisotropic interaction and collapse-revival dynamics in the AQRM}.
			\textbf{a,b,} Measured time-resolved qubit excitation probability $P_g$ and cavity photon population $n$ as functions of the anisotropy ratio $\lambda$.
			\textbf{c,} Extracted total photon number $N$ for different values of $\lambda$.
			\textbf{d,} Evolution of the qubit populations for $\lambda=5$ AQRM from the dual mapping, by applying $X$ and $X/2$ gates to the system qubit respectively. The parameters are $\Omega/2\pi=\omega/2\pi=1$~MHz and $g_0/2\pi=2$~MHz.
			\textbf{e--g,} Collapse-revival dynamics for different values of $\lambda$ in the degenerate case. The cases $\lambda=1$, $0.8$, and $0.6$ are shown in blue, orange, and green, respectively. The parameters are $\omega/2\pi=1.6$~MHz and $g_0/2\pi=4.5$~MHz, corresponding to $\min\{g_1/\omega,\, g_2/\omega\}=1.4$, $1.25$, and $1.05$, respectively, which are all well within the DSC regime.
			\textbf{h,} Adjacent level spacings $(E_{j+1}-E_j)$ in the positive- and negative-parity sectors. \textbf{i,} Photon-number fluctuation $\delta n$ extracted from the dynamics for different $\lambda$. Symbols represent experimental data, and dashed curves show numerical results from the effective AQRM. Solid curves include the full experimental protocol, decoherence, and the multilevel structure of the transmon treated as a qutrit.
		}
		\label{fig2}
	\end{figure*}
	
	\begin{figure*}
		\centering
		\includegraphics[width=0.99\linewidth]{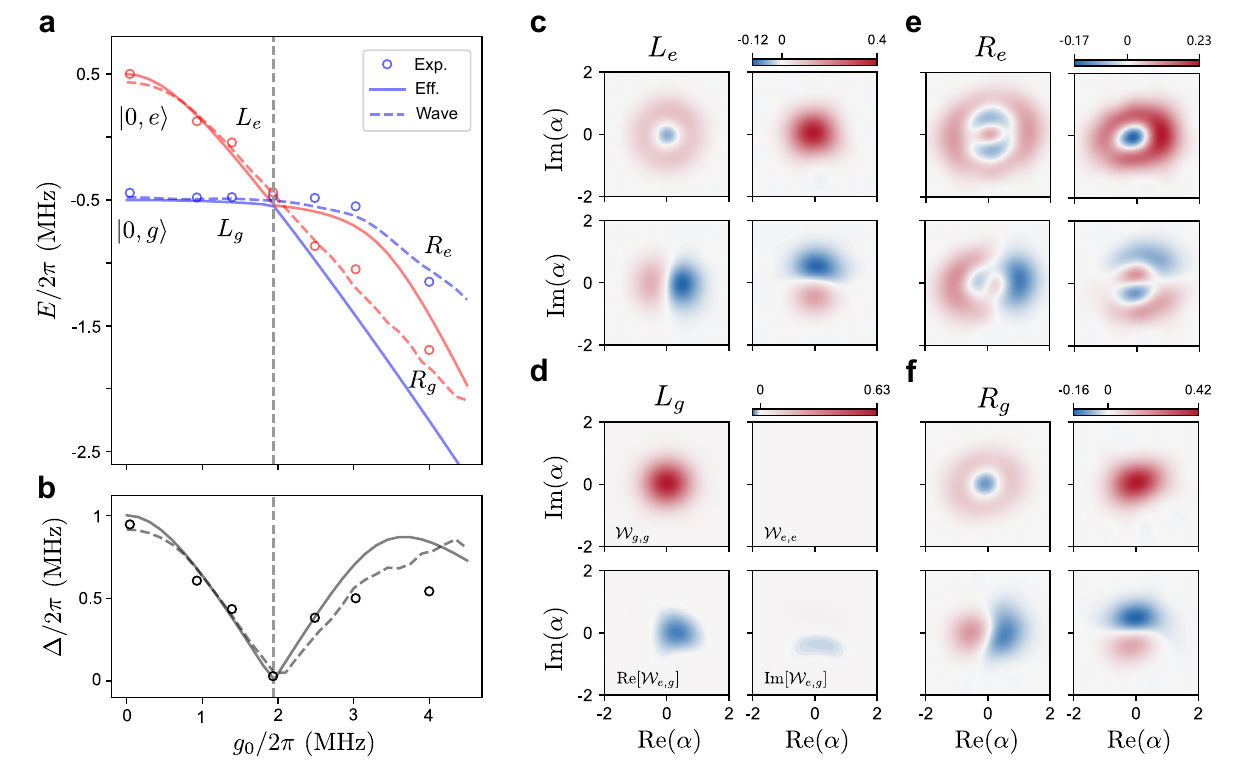}
		\caption{
			\textbf{Ground-state level crossing and phase-space signatures of the AQRM.}
			\textbf{a,} Energy spectrum as a function of $g_0=g_1+g_2$, showing an exact crossing between the lowest even- and odd-parity branches. The solid curves correspond to numerical results calculated via the effective Hamiltonian, whereas the dashed curves are obtained using the same wave protocol adopted in the experiment. \textbf{b,} Energy gap between the ground state and the first excited state. The dashed line denotes the predicted crossing point $g_{0c}$.
			\textbf{c--f,} Joint Wigner function matrix at $g_0/2\pi=1.5$ ($4$)~MHz for $\ket{0, g},\ket{0, e}$ as initial states, labelled as $L_g (R_g)$ and $L_e (R_e)$ individually. The parameters are $\Omega/2\pi=1$~MHz, $\omega/2\pi=2.5$~MHz, and $\lambda=0.2$. The colorbars are set white when the value is zero, red when positive, and blue as negative.
		}
		\label{fig3}
	\end{figure*}
	
	Here, we construct a programmable AQRM simulator based on the superconducting circuits, with independent control of the rotating and counterrotating channels $(g_1,g_2)$ and of the bias $\varepsilon$. Together with a duality mapping, this tunability allows us to access the anisotropy continuously from $\lambda=0$ to $\lambda=\infty$, thereby covering the full AQRM parameter space. Using the time-domain measurements, adiabatic state engineering and joint qubit-resonator tomography, we show how anisotropy and bias reshape the QRM in three connected ways. First, anisotropy modifies the level spacings in the DSC regime and converts complete collapse-revival dynamics into incomplete revivals. Second, it reorganizes the low-lying spectrum and produces a ground-state parity switch. Third, for the biased models, it shifts the dynamical signature of hidden symmetry in a way that follows the predicted anisotropic commensurability condition. Together these observations turn the AQRM from a theoretical interpolation into an experimentally programmable symmetry landscape.
	
	\subheading{Experimental setup}
	Our experiment is performed on a flip-chip superconducting quantum processor, comprising a central bus resonator $R$ and 24 frequency-tunable transmon qubits~\cite{supp_cite}. We choose one transmon as the system qubit $Q$, coupled to $R$ through a tunable coupler $C$ for the in-situ control of the qubit--resonator interaction~\cite{Yan2018}. The intracavity photon state is measured using an ancillary transmon $Q_a$ dispersively coupled to $R$, enabling photon-number-resolved detection~\cite{LinPeng2013}. The system qubit has an anharmonicity $\alpha/2\pi=-0.37~\mathrm{GHz}$, which is sufficiently large to suppress leakage and support an effective two-level description.
	
	The central capability of the experiment is to convert microwave control into the independent control of the two quadratures of the light--matter interaction. We engineer the AQRM by combining the longitudinal qubit frequency modulation $\tilde{\omega}_q$ with a multi-tone transverse drive $\mathcal{E}$. The laboratory-frame Hamiltonian is
	\begin{equation}
		\hat{H}_{\textrm{lab}}=
		\frac{\tilde{\omega}_q}{2}\hat{\sigma}_z+\omega\hat{a}^{\dagger}\hat{a}
		+\left(g_0\hat{a}^{\dagger}\hat{\sigma}_-+\mathcal{E}\hat{\sigma}_-+\mathrm{H.c.}\right),
	\end{equation}
	where $\tilde{\omega}_q=\omega_q+A\cos(\mu t)$ and $\mathcal{E}=[V_0+V\cos(\nu t)]e^{i\omega_0t}$. The mechanism can be understood from a spin-$1/2$ picture. The cavity field acts as a transverse magnetic field with two quadratures,
	$B^x=g_0(\hat a+\hat a^\dagger)/2$ and
	$B^y=ig_0(\hat a^\dagger-\hat a)/2$~\cite{Zheng2023}. In the rotating frame with $\omega_0$, the carrier tone sets the bias as $\varepsilon=2V_0$, while the transverse modulation produces an $x$-field, $B_0^x(t)=V\cos(\nu t)$. The rapid modulation of the total $x$ field, $B^x+B_0^x(t)$, renormalizes the $B^y$ component with the Bessel factor $J_0(2\beta)$, with $\beta=V/\nu$, because $\hat{\sigma}_y$ couples eigenstates of $\hat{\sigma}_x$. This mechanism is analogous to the dynamical control of hopping in the driven lattice systems~\cite{Lignier2007,Shi2023,Liu2025}.
	Consequently, the two cavity quadratures acquire different effective strengths, and the stroboscopic dynamics are described by the AQRM Hamiltonian. The total coupling $g_0$ is partitioned into rotating and counterrotating components $g_1$ and $g_2$, with $g_1+g_2=g_0$, and the anisotropy ratio becomes
	$\lambda=(1-J_0(2\beta))/(1+J_0(2\beta))$.
	This allows $\lambda$ to be continuously tuned from $0$ to $2.3$ in our implementation. Because the effective USC/DSC coupling is generated with Floquet synthesis rather than through an intrinsically bare coupling, this approach avoids the equilibrium constraints associated with no-go arguments~\cite{Nataf2010,Ballester2012}.
	
	We first verify that this control does more than rescale the interaction strength: it changes the symmetry of the dynamics. To benchmark the engineered interaction, we isolate
	$\hat{H}_{\mathrm{int}}=g_1(\hat{H}_{\mathrm{r}}+\lambda\hat{H}_{\mathrm{cr}})$
	and initialize the system in a single-photon state $\ket{0,e}$. Figures.~\ref{fig2}\textbf{a} and \textbf{b} show the measured qubit and resonator populations, as $\lambda$ is swept at fixed $g_0/2\pi=3~\mathrm{MHz}$. We further show the extracted total photon number in Fig.~\ref{fig2}\textbf{c}, which provides a compact probe of $U(1)$-symmetry breaking and multiphoton generation. For $\lambda=0$, $N$ remains conserved, as expected from the JCM symmetry. Increasing $\lambda$ activates counterrotating processes, mixes excitation manifolds, and drives the system from bounded JCM oscillations toward the photon growth of the isotropic point $\lambda=1$, where $N\simeq g^2t^2$ in the DSC limit~\cite{Monroe1996,supp_cite}. For $\lambda>1$, the dynamics are suppressed, because $\hat H_{\rm cr}$ does not act on $\ket{0,e}$ at first order. The extracted photon number captures this symmetry-controlled crossover and agrees well with theoretical prediction.
	
	The rotating and counterrotating processes are further connected by introducing the identity $\hat{H}(\omega,\Omega,g_1,g_2,\varepsilon)=\hat{\sigma}_x\hat{H}(\omega,-\Omega,g_2,g_1,\varepsilon)\hat{\sigma}_x$,
	which maps between $0<\lambda<1$ and $\lambda>1$ sectors~\cite{supp_cite}. Experimentally, this duality allows a target AQRM with anisotropy $\lambda$ to be realized by implementing its dual Hamiltonian at $1/\lambda$ with opposite qubit frequency, together with $\pi$ pulses before and after the evolution. Using this scheme, we realize the dynamics at $\lambda=5$ through the dual point $\lambda=0.2$, for the initial states $\ket{0,e}$ and $(\ket{0,g}-i\ket{0,e})/\sqrt{2}$, as shown in Fig.~\ref{fig2}\textbf{d}. The agreement with the target dynamics demonstrates the access to the full AQRM parameter space, spanning the JCM, QRM, and anti-JCM limits~\cite{BocanegraGaray2024}.
	
	\begin{figure*}[!t]
		\centering
		\includegraphics[width=0.99\linewidth]{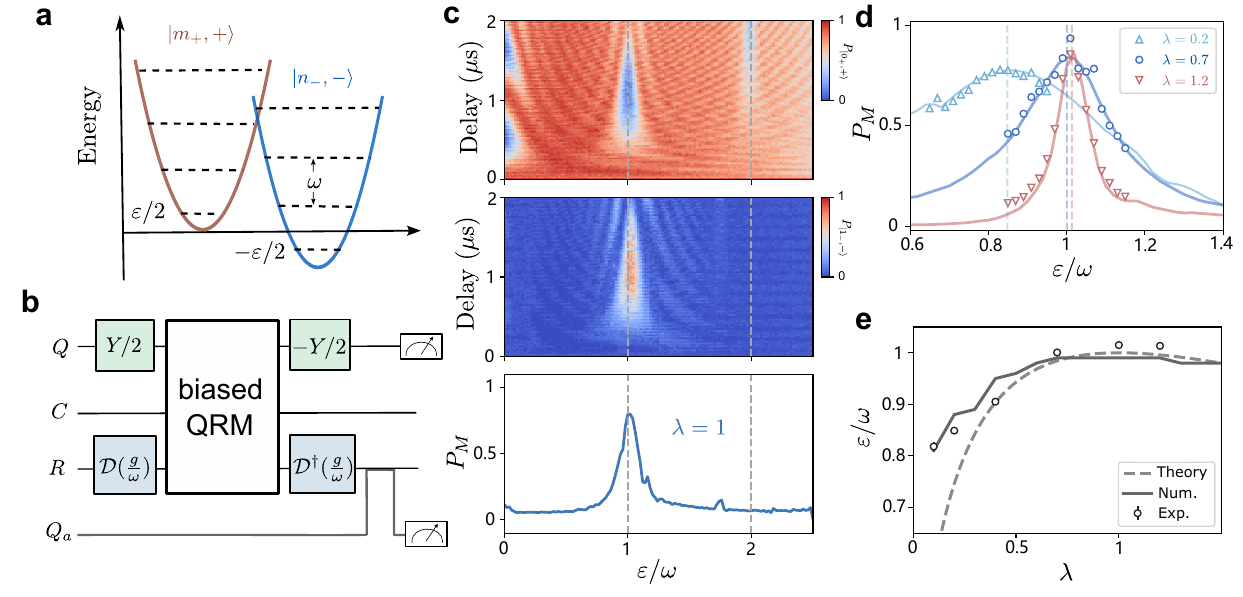}
		\caption{\textbf{Hidden symmetry and tunnelling dynamics in biased models.}
			\textbf{a}, Schematic effective potentials of the biased QRM for $\varepsilon\neq 0$, where the $\mathbb{Z}_2$ symmetry is broken.
			\textbf{b}, Pulse sequence for observing tunnelling dynamics in $P_{\ket{0_+,+}}$ in the biased QRM by mapping bare number states onto the displaced Fock states.
			\textbf{c}, The time evolution of QRM starting from the initial state $\ket{0_+,+}$ against $\varepsilon/\omega$. The upper, middle, and lower panels show $P_{\ket{0_+,+}}$, $P_{\ket{1_-,-}}$, and the extracted $P_M$, respectively. The dashed lines denote $\varepsilon/\omega=0, 1, 2$.
			\textbf{d}, The transfer population $P_M$ as a function of $\varepsilon/\omega$ for several values of $\lambda$. The dashed lines represent the position where $P_M$ achieves a peak, where the solid curves denotes the numerical results.
			\textbf{e}, Extracted position of $P_M$ peaks as a function of $\lambda$.  The dashed curve denotes the theoretical hidden symmetry points, while the solid curve represent the numerical results using the same protocol and parameters as the experiments. Error bars represent one standard deviation.}
		\label{fig4}
	\end{figure*}
	
	\subheading{Collapse-revival dynamics in the DSC regime}
	Collapse-revival dynamics provides a direct time-domain probe of the spectral regularity in the USC/DSC regime~\cite{Eberly1980,Casanova2010} and has been widely explored in the QRM across diverse platforms~\cite{Braumuller2017, Langford2017, Lv2018, Koch2023}. In Fig.~\ref{fig2}\textbf{e}, we present the DSC dynamics of the QRM at $g/\omega=1.4$ in the degenerate case $\Omega=0$, initialized from the JCM ground state $\ket{0, g}$. The initial-state population almost decays to zero, remains suppressed over a period of $2\pi/\omega$, and then revives completely. This behaviour can be understood from a displacement transformation $\mathcal{D}(g/\omega)$~\cite{Casanova2010}, with $\mathcal{D}(\eta)=e^{\eta\hat{a}^{\dagger}-\eta^*\hat{a}}$. In the degenerate isotropic QRM, this transformation maps the system to an effective harmonic oscillator with equal level spacing $\omega$, so the dynamical phases rephase exactly after one period. For $\lambda\neq 1$, however, the eigenenergies are roots of a transcendental function~\cite{Braak2011,Xie2014} and are no longer equally spaced, interrupting the interference that produces perfect revivals.
	
	In Figs.~\ref{fig2}\textbf{f} and \textbf{g}, we present the DSC dynamics for the anisotropic cases for $\lambda=0.8$ and $0.6$, respectively, using the same $\Omega$, $\omega$, and $g_0$ as in the isotropic case. The initial-state population still exhibits a collapse-revival pattern, but the revival is incomplete, which is consistent with the breakdown of the equally spaced spectral structure.  In Fig.~\ref{fig2}\textbf{h}, we show the corresponding level spacings $E_{j+1}-E_j$ in the two parity subspaces. We further characterize the dynamics using the photon-number fluctuation $\delta n=\sqrt{\langle \hat{n}^2\rangle-\langle \hat{n}\rangle^2}$, as shown in Fig.~\ref{fig2}\textbf{i}. For $\lambda=1$, the state returns to $\ket{0,g}$ up to a global phase at integer multiples of $2\pi/\omega$~\cite{Casanova2010}, yielding $\delta n=0$ at those instants. For $\lambda=0.6$ and $0.8$, the minima of $\delta n$ are lifted, showing that anisotropy leaves a dynamical fingerprint even in the nominally degenerate limit. These results connect a programmable microscopic imbalance directly to a measurable reconstruction of the spectral interference.
	
	\subheading{Ground-state parity switching}
	Spectral reconstruction is also visible in the low-lying eigenstates. For $\varepsilon=0$, the AQRM preserves the same $\mathbb{Z}_2$ parity symmetry as the QRM, splitting the spectrum into even- and odd-parity branches. The generic branches form the regular spectrum, while isolated crossings between opposite-parity branches occur only at the special parameter values, giving rise to the exceptional spectrum. A hallmark of anisotropy is an exact crossing between the two lowest levels of opposite parity. As $g_0$ increases at a fixed $\lambda$, this crossing exchanges the ground and first-excited states, producing a ground-state parity switch absent in the isotropic QRM. For the first crossing with $0\le\lambda<1$, the crossing point is given by the Judd-type exceptional solution $g_{0c}=(1+\lambda)g_{1c}$~\cite{Xie2014}, where $g_{1c}=\sqrt{\Omega\omega/(1-\lambda^2)}$. This provides a sharp spectral landmark of the AQRM.
	
	To locate this crossing experimentally, we track the two lowest energies using adiabatic state preparation followed by joint qubit-resonator tomography~\cite{LinPeng2013}. Starting from the decoupled point, the qubit is initialized in $\ket{g}$ or $\ket{e}$ and the coupler is adiabatically ramped to the target interaction strength, mapping the product states onto low-lying eigenstates of the coupled Hamiltonian. Tomography yields the joint density matrix $\rho$ and the energy expectation $\mathrm{Tr}[\rho\hat{H}]$, allowing the two branches and their interchange to be resolved. In Figs.~\ref{fig3}\textbf{a} and \textbf{b}, we measure the ground and first-excited energies as a function of $g_0$ and identify an exact crossing at the predicted $g_{0c}$.
	
	The joint qubit-resonator state can be fully characterized by a Wigner function matrix~\cite{Lutterbach1997,Bertet2002,LinPeng2013,Vlastakis2013,Kirchmair2013,Zheng2023}, with elements $\mathcal{W}_{kk'}(\alpha)$ constructed from the cavity blocks $\rho_{kk'}=\langle k|\rho|k'\rangle$ for $k,k'\in\{g,e\}$. Since the Wigner value at the origin gives the photon-number parity $\mathcal{W}(0)=2\langle(-1)^{\hat n}\rangle/\pi$~\cite{Lutterbach1997,Bertet2002,Hofheinz2009}, the total parity is obtained from the qubit-resolved components as
	\begin{equation}
		\mathcal{P}=\frac{\pi}{2}\big[\mathcal{W}_{gg}(0)-\mathcal{W}_{ee}(0)\big].
	\end{equation}
	The measured Wigner function matrices in Fig.~\ref{fig3}\textbf{c}-\textbf{f} show that the ground state changes from even parity ($\mathcal{P}=1.02$ for $g_0<g_{0c}$) to odd parity ($\mathcal{P}=-0.91$ for $g_0>g_{0c}$). This directly reveals an anisotropy-induced parity switch from the competition between rotating and counterrotating channels.
	
	\subheading{Hidden symmetry and tunnelling dynamics}
	A final test of the programmable symmetry is provided by the biased model. Although a nonzero bias breaks the $\mathbb{Z}_2$ parity symmetry and generally lifts parity-protected degeneracies, exact level crossings can reemerge at discrete commensurate biases owing to hidden symmetries. At these points, wave-packet dynamics are confined to a reduced subspace spanned by displaced polaron states, enabling coherent tunnelling between quasi-degenerate effective potential wells~\cite{Li2021}, as illustrated in Fig.~\ref{fig4}\textbf{a}. For the QRM, the hidden symmetry emerges when the bias satisfies the commensurate condition
	\begin{equation}
		\varepsilon_{h}/\omega = k,
		\label{eq_tunneling_qrm}
	\end{equation}
	where $k$ is a positive integer. Under this resonance condition, the tunnelling is mediated by states that are direct tensor products of displaced Fock states and transversely polarized qubit states,
	$|n_\pm,\pm\rangle=\mathcal{D}(\pm g/\omega)\ket{n}\otimes\ket{\pm}$,
 	where $\ket{\pm}=(\ket{e}\pm\ket{g})/\sqrt{2}$. Because this tunnelling is a high-order, symmetry-selected resonance with a narrow linewidth, it is sensitive to parameter fluctuations and pulse imperfections, making its observation a stringent test of the simulator control accuracy. Experimentally, we align the evolution, initial state, and measurement basis with the hidden-symmetry subspace using joint displacement operations and single-qubit gates before and after the biased QRM's evolution, as shown in Fig.~\ref{fig4}\textbf{b}. These operations map the displaced Fock-state basis onto the bare photon-number basis for measurement. Figure.~\ref{fig4}\textbf{c} shows the dynamics of $P_{\ket{0_+,+}}$ and $P_{\ket{1_-,-}}$ as a function of $\varepsilon/\omega$, initialized from $\ket{0_+,+}$. In the lower panel, we plot the extracted maximum transfer probability, $P_M=\max_t P_{\ket{1_-,-}}(t)$, as a function of $\varepsilon/\omega$. The tunnelling points originated from hidden symmetries are clearly identified near $\varepsilon/\omega=1.0$, corresponding to the commensurate biases in Eq.~\eqref{eq_tunneling_qrm}.
	
	For $\lambda\neq1$, the hidden-symmetry points become $\lambda$ dependent and are given by~\cite{Li2021}
	\begin{equation}
		\varepsilon_h/\omega
		=
		k\frac{2\sqrt{\lambda}}{1+\lambda},
		\label{eq_tunneling_aqrm}
	\end{equation}
	which reduces to Eq.~\eqref{eq_tunneling_qrm} at $\lambda=1$. Although an analogous tunnelling-dynamics picture can be intuitively constructed, obtaining an analytic form of the corresponding tunnelling initial state is difficult in the anisotropic case. We therefore prepare a $\lambda$-dressed initial state, $\ket{0_+,+}_{\lambda}=\mathcal{D}(-\sqrt{\lambda}g_1/\omega)\ket{0}\otimes
	(\sqrt{\lambda}\ket{g}+\ket{e})/\sqrt{1+\lambda}$ and measure the population at $\ket{1_-,-}_{\lambda}=\mathcal{D}(\sqrt{\lambda}g_1/\omega)\ket{1}\otimes
	(\ket{g}-\sqrt{\lambda}\ket{e})/\sqrt{1+\lambda}$,
	which can be effectively derived in the weak-coupling regime~\cite{supp_cite}. Figure~\ref{fig4}\textbf{d} shows $P_M$ as a function of $\varepsilon/\omega$, revealing a $\lambda$-dependent shift of the peak positions. In Fig.~\ref{fig4}\textbf{e}, we extract the value of $\varepsilon/\omega$ at which $P_M$ reaches its maximum and plot it as a function of $\lambda$. The extracted trend follows the predicted anisotropy-dependent displacement of the hidden-symmetry point in Eq.~\eqref{eq_tunneling_aqrm}, providing a dynamical signature of generalized hidden symmetry in the AQRM.
	
	\subheading{Conclusion and outlook}
	In summary, we have realized a programmable extension of the quantum Rabi model in which anisotropy and bias become experimentally tunable axes of symmetry control. This capability goes beyond reaching a large effective coupling: it allows the balance between energy-conserving and nonconserving processes to be engineered, thereby controlling spectral structure, protected crossings and dynamical selection rules. The observed incomplete revivals, ground-state parity switching and hidden-symmetry tunnelling provide complementary views of the same underlying principle: the nonperturbative light--matter dynamics can be shaped by programming the symmetry of the Hamiltonian. Looking forward, the approach can be extended to time-dependent anisotropic protocols, multi-qubit anisotropic Dicke-type models~\cite{Baksic2014,Buijsman2017}, and more complex Floquet-engineered light-matter Hamiltonians~\cite{Akbari2025}.

	\clearpage
	\newpage
	
	\bibliography{main}
	
	~\\
	\subheading{Code availability}
	The codes used for the numerical simulations are available from the corresponding authors upon request.
	
	~\\
	\subheading{Acknowledgements}
	We thank R.-H. Zheng for helpful discussions. We thank the support from the Synergetic Extreme Condition User Facility (SECUF) in Huairou District, Beijing. Devices were made at the Nanofabrication Facilities at Institute of Physics, CAS in Beijing. This work was supported by the National Natural Science Foundation of China (Grants No. 92265207, T2121001, T2322030, 12122504, 12274142, 12475017, 12504593, 92365206, 12104055, 12304390, 12574386, and U25A6009), Scientific Research Innovation Capability Support Project for Young Faculty (Grant No. SRICSPYF-ZY2025171), the Natural Science Foundation of Guangdong Province (Grant No. 2024A1515010398), the Guangdong Provincial Quantum Science Strategic Initiative (Grant No. GDZX2505004), the Startup Grant of South China University of Technology (SCUT) (Grant No. 20240061),  the National Postdoctoral Overseas Talent Recruitment Program of China, the Fujian 100 Talents Program, the Fujian Minjiang Scholar Program, the Innovation Program for Quantum Science and Technology (Grant No. 2021ZD0301800), the Beijing Nova Program (No. 20220484121), and the China Postdoctoral Science Foundation (Grants No. GZC20252227, and 2025M783453).
    
    ~\\
	\subheading{Competing interests}
	The authors declare no competing interests.
	
\end{document}


\title{\textit{Supplementary Information for:}\\ ``Programmable spectral symmetries in an anisotropic quantum Rabi simulator''}
	
		\author{Jia-Cheng Song}
	\thanks{These authors contributed equally to this work.}
	\affiliation{\addrA}
	\affiliation{\addrB}
	
	\author{Yu Liu}
	\email{yuliu@iphy.ac.cn}
	\thanks{These authors contributed equally to this work.}
	\affiliation{\addrA}
    
	\author{Ming-Chuan Wang}
	\thanks{These authors contributed equally to this work.}
	\affiliation{\addrA}
	\affiliation{\addrB}
	
	\author{Ke-Xiong Yan}
	\thanks{These authors contributed equally to this work.}
	\affiliation{\addrF}
	
	\author{Yang He}
	\affiliation{\addrA}
	\affiliation{\addrB}
	
	\author{Yun-Hao Shi}
	\affiliation{School of Physics, Xi’an Jiaotong University, Xi’an 710049, China}
	
	\author{Wei-Ping Yuan}
	\affiliation{\addrA}
	\affiliation{\addrB}
	
	\author{Cheng-Lin Deng}
	\affiliation{\addrC}
	
	\author{Li Li}
	\affiliation{\addrA}
	\affiliation{\addrB}
	
	\author{Zhen-Ting Bao}
	\affiliation{\addrA}
	\affiliation{\addrB}
	
	\author{Yutao Chen}
	\affiliation{\addrA}
	\affiliation{\addrB}
	
	\author{Xu-Yang Gu}
	\affiliation{\addrA}
	\affiliation{\addrB}
	
	\author{Tian-Ming Li}
	\affiliation{\addrA}
	\affiliation{\addrB}
	
	\author{Gui-Han Liang}
	\affiliation{\addrA}
	
	\author{Zheng-He Liu}
	\affiliation{\addrA}
	\affiliation{\addrB}
	
	\author{Wei-Guo Ma}
	\affiliation{\addrA}
	\affiliation{\addrB}
	
	\author{Zhen-Yu Peng}
	\affiliation{\addrA}
	\affiliation{\addrB}
	
	\author{Shuai-Li Wang}
	\affiliation{\addrA}
	\affiliation{\addrB}
	
	\author{Yong-Xi Xiao}
	\affiliation{\addrA}
	\affiliation{\addrB}
	
	\author{Yi-Han Yu}
	\affiliation{\addrA}
	\affiliation{\addrB}
	
	\author{Jia-Chi Zhang}
	\affiliation{\addrA}
	\affiliation{\addrB}
	
	\author{Kui Zhao}
	\affiliation{\addrC}
	
	\author{Min-Xuan Zhou}
	\affiliation{\addrA}
	\affiliation{\addrB}
	
	\author{Kaixuan Huang}
	\affiliation{\addrC}
	
	\author{Yu-Ran Zhang}
	\affiliation{School of Physics and Optoelectronics, South China University of Technology, Guangzhou 510640, China}
	
	\author{Yu-Xiang Zhang}
	\affiliation{\addrA}
	\affiliation{\addrB}
	\affiliation{\addrD}
	
	\author{Zhongcheng Xiang}
	\affiliation{\addrA}
	\affiliation{\addrB}
	\affiliation{\addrD}

    \author{Dongning Zheng}
	\affiliation{\addrA}
	\affiliation{\addrB}
	\affiliation{\addrD}
	\affiliation{\addrE}
	
	\author{Ye-Hong Chen}
	\email{yehong.chen@fzu.edu.cn}
	\affiliation{\addrF}
	\affiliation{Quantum Information Physics Theory Research Team, Center for Quantum Computing, RIKEN, Wako-shi, Saitama 351-0198, Japan}
	
	\author{Kai Xu}
	\email{kaixu@iphy.ac.cn}
	\affiliation{\addrA}
	\affiliation{\addrB}
	\affiliation{\addrC}
	\affiliation{\addrD}
	
	\author{Heng Fan}
	\email{hfan@iphy.ac.cn}
	\affiliation{\addrA}
	\affiliation{\addrB}
	\affiliation{\addrC}
	\affiliation{\addrD}
	\affiliation{\addrE}

	\date{\today}
	\maketitle
	\tableofcontents
	\newpage
	
	\section{Model and Hamiltonian}
	
	\subsection{Anisotropic quantum Rabi model}

	The anisotropic quantum Rabi model (AQRM)~\cite{Xie2014} generalizes the standard quantum Rabi model (QRM) by allowing the rotating and counterrotating interactions to be controlled by two independent coupling constants. The Hamiltonian reads ($\hbar=1$)
    \begin{equation}\label{eqAQRM}
        \hat{H}=\omega\hat{a}^{\dagger}\hat{a}+\frac{\Omega}{2}\hat{\sigma}_z+g_1(\hat{a}^{\dagger}\hat{\sigma}_-+\hat{a}\hat{\sigma}_+)+g_2(\hat{a}^{\dagger}\hat{\sigma}_++\hat{a}\hat{\sigma}_-)+\frac{\varepsilon}{2}\hat{\sigma}_x,
    \end{equation}
    where $\hat{a}^{\dagger}$ ($\hat{a}$) is the creation (annihilation) operator, $\hat{\sigma}_{x,y,z}$ are the Pauli matrices, $\hat{\sigma}_{\pm}=(\hat{\sigma}_x\pm \mathrm{i}\hat{\sigma}_y)/2$, $\omega$ is the bosonic-mode frequency, $\Omega$ is the qubit transition frequency, $g_1$ and $g_2$ are the coupling strengths of the rotating and counterrotating terms, respectively, and $\varepsilon$ is the qubit bias. By introducing the anisotropy ratio $\lambda=g_2/g_1$, the AQRM interpolates between the Jaynes-Cummings model (JCM) ($\lambda=0$) and the QRM ($\lambda=1$). For the JCM with $\varepsilon=0$, the total excitation number is conserved:
    \begin{equation}
        \hat{N}=\hat{a}^{\dagger}\hat{a}+\frac{1}{2}(1+\hat{\sigma}_z).
    \end{equation}
    For $\lambda\neq 0$ and $\varepsilon=0$, the $U(1)$ symmetry is broken to a discrete $\mathbb{Z}_2$ parity symmetry, i.e., $[\hat{H},\hat{\mathcal{P}}]=0$, with the parity operator
    \begin{equation}
        \hat{\mathcal{P}}=\mathrm{e}^{\mathrm{i}\pi\hat{N}}=-(-1)^{\hat{a}^{\dagger}\hat{a}}\hat{\sigma}_z.
    \end{equation}
    In contrast, when $\varepsilon\neq 0$, parity symmetry is broken and the energy degeneracies are reorganized. Nevertheless, a hidden symmetry can restore level crossings at special values of $\varepsilon$, a feature that has attracted considerable recent interest~\cite{Li2021, Mangazeev2021, ReyesBustos2021}.
    
    The AQRM admits an alternative formulation that reveals its interpretation as a two-level system coupled to both quadratures of a bosonic mode. This becomes evident by rewriting
	\begin{equation}
	     \hat{H}=\omega\hat{a}^{\dagger}\hat{a}+\frac{\Omega}{2}\hat{\sigma}_z+g_x(\hat{a}+\hat{a}^{\dagger})\hat{\sigma}_x+\mathrm{i}g_y(\hat{a}-\hat{a}^{\dagger})\hat{\sigma}_y+\frac{\varepsilon}{2}\hat{\sigma}_x,
	\end{equation}
	where $g_x=(g_1+g_2)/2$ and $g_y=(g_1-g_2)/2$. The model can also be mapped onto a two-dimensional electron gas with both Rashba and Dresselhaus spin-orbit couplings under a perpendicular magnetic field. Here we realize the AQRM in a circuit-QED architecture.

	
	\subsubsection{Derivation of the effective AQRM}
	Our approach builds on the method originally developed in Ref.~\cite{Ballester2012}. We start from the JCM,
	\begin{subequations}
	\begin{align}
	    \hat{H}_0=&\frac{\omega_q}{2}\hat{\sigma}_z+\omega_r\hat{a}^{\dagger}\hat{a}+g_0(\hat{a}^{\dagger}\hat{\sigma}_-+\hat{a}\hat{\sigma}_+)\\
	    =&\frac{\omega_q}{2}\hat{\sigma}_z+\omega_r\hat{a}^{\dagger}\hat{a}+\frac{g_0}{2}(X\hat{\sigma}_x+Y\hat{\sigma}_y),
	\end{align}
	\end{subequations}
	where $X=(\hat{a}+\hat{a}^{\dagger})$ and $Y=i(\hat{a}-\hat{a}^{\dagger})$. This Hamiltonian provides an excellent description of standard circuit quantum electrodynamics (cQED) devices~\cite{FornDaz2019, Blais2021}. Both the qubit frequency $\omega_q$ and the coupling strength $g_0$ can be finely tuned~\cite{Yan2018}. The key step in synthesizing the AQRM is to introduce a three-tone transverse field together with a longitudinal drive. The corresponding Hamiltonian in the laboratory frame is
	\begin{equation}\label{ham_driven}
		\hat{H}=\frac{\omega_q}{2}\hat{\sigma}_z+\frac{A}{2}\cos\mu t\hat{\sigma}_z+\omega_r\hat{a}^{\dagger}\hat{a}+g_0(\hat{a}^{\dagger}\hat{\sigma}_-+\hat{a}\hat{\sigma}_+)+[V_0+V\cos(\nu t)](\mm{e}^{\mm{i}\omega_0t}\hat{\sigma}_-+\mm{H.c.}),
	\end{equation} 
	where $A$ and $\mu$ are the amplitude and frequency of the longitudinal field, respectively; $\{V_0, V/2\}$ are the amplitudes of the three-tone transverse field; and $\{\omega_0,\omega_0\pm\nu\}$ are the corresponding drive frequencies. We first perform the unitary transformation
	\begin{equation}\label{U1}
		\hat{U}_1=\exp\left\{i\omega_0t(\frac{1}{2}\hat{\sigma}_z+\hat{a}^{\dagger}\hat{a})\right\},
	\end{equation}
	yielding the system Hamiltonian in the rotating frame:
	\begin{subequations}\label{eqHamAfterU1}
		\begin{align}
			\hat{H}_I=&\hat{U}\hat{H}\hat{U}^{\dagger}-i\hat{U}\dot{\hat{U}}^{\dagger}\\
			=&\frac{1}{2}(\omega_q-\omega_0+A\cos\mu t)\hat{\sigma}_z+(\omega_r-\omega_0)\hat{a}^{\dagger}\hat{a}+g_0(\hat{a}^{\dagger}\hat{\sigma}_-+\hat{a}\hat{\sigma}_+)+[V_0+V\cos(\nu t)]\hat{\sigma}_x.
		\end{align}
	\end{subequations}
	The above equation can be understood in a spin-$1/2$ picture, by rewriting it as
	\begin{equation}
	    \hat{H}_I=(\frac{g_0}{2}X+V_0+V\cos(\nu t))\hat{\sigma}_x+\frac{g_0}{2}Y\hat{\sigma}_y+\frac{1}{2}(\omega_q-\omega_0+A\cos\mu t)\hat{\sigma}_z+(\omega_r-\omega_0)\hat{a}^{\dagger}\hat{a}.
	\end{equation}
	The spin is subjected to a transverse magnetic field induced by a quantized cavity field. Since $\hat{\sigma}_y$ induces tunneling between the eigenstates of $\hat{\sigma}_x$, fast modulation of the $\hat{\sigma}_x$ term can renormalize the $\hat{\sigma}_y$ component, in analogy with the dynamical control of hopping in driven lattice systems~\cite{Lignier2007, Shi2023, Liu2025}.
	
	\begin{figure}[t]
	    \centering
	    \includegraphics[width=0.5\linewidth]{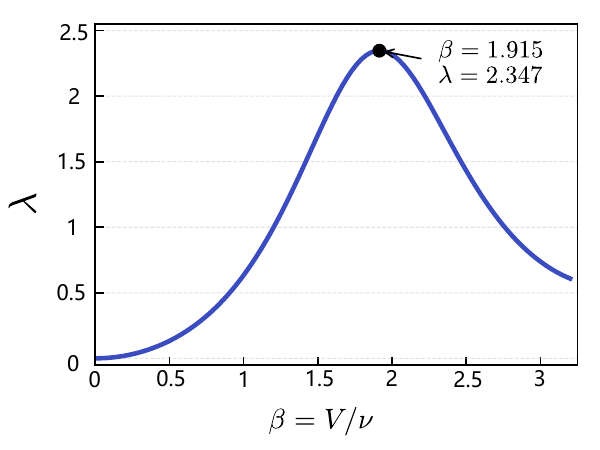}
	    \caption{Effective tunable range of $\lambda$ as a function of $\beta=V/\nu$.}
	    \label{tunable_range}
	\end{figure}
	
	Specifically, we implement a second unitary transformation
	\begin{equation}\label{U2}
	    \hat{U}_2=\exp\left\{\mm{i}\beta\sin(\nu t)\hat{\sigma}_x\right\},
	\end{equation}
	with $\beta=V/\nu$. The system Hamiltonian in this new frame can be calculated as
	\begin{subequations}\label{eqHamU2}
		\begin{align}
			\hat{H}_I'=&\mm{e}^{\mm{i}\beta\sin\nu t\hat{\sigma}_x}[\hat{H}_I-V\cos\nu t\hat{\sigma}_x]\mm{e}^{-\mm{i}\beta\sin\nu t\hat{\sigma}_x}\\
			=&(\omega_r-\omega_0)\hat{a}^{\dagger}\hat{a}+\frac{1}{2}(\omega_q-\omega_0+A\cos\mu t)
			\begin{bmatrix}
				\cos (2\beta\sin\nu t)&\mm{i}\sin (2\beta\sin \nu t)\\
				-\mm{i}\sin (2\beta\sin \nu t)&-\cos (2\beta\sin\nu t)
			\end{bmatrix}\\
			&+\frac{g_0}{2}(\hat{a}^{\dagger}
			\begin{bmatrix}
				-\mm{i}\sin (2\beta\sin\nu t)&1-\cos (2\beta\sin\nu t)\\
				1+\cos (2\beta\sin\nu t)&\mm{i}\sin (2\beta\sin\nu t)
			\end{bmatrix}
		+\mm{H.c.})+V_0\hat{\sigma}_x.
		\end{align}
	\end{subequations}
	The qubit-frequency term becomes
	\begin{equation}
	    \hat{H}'_q=\frac{1}{2}(\omega_q-\omega_0+A\cos\mu t)[\cos(2\beta\sin\nu t)\hat{\sigma}_z-\sin(2\beta\sin\nu t)\hat{\sigma}_y],
	\end{equation}
	and the interaction term can be written as
	\begin{equation}
	    \hat{H}'_{\textrm{int}}=\frac{g_0}{2}(\hat{a}+\hat{a}^{\dagger})\hat{\sigma}_x+\mathrm{i}\frac{g_0}{2}\cos(2\beta\sin\nu t)(\hat{a}-\hat{a}^{\dagger})\hat{\sigma}_y.
	\end{equation}
	Using the Jacobi-Anger expansion
	\begin{equation}
		e^{\mm{i}\alpha\sin\nu t}=\sum_{n=-\infty}^{\infty}J_n(\alpha)e^{\mm{i}n\nu t},
	\end{equation}
	where $J_n(\cdot)$ is the $n$-th Bessel function of the first kind. This gives
	\begin{subequations}
		\begin{align}
			\cos(2\beta\sin\nu t)=&J_0(2\beta)+2\sum_{k=1}^{\infty}J_{2k}(2\beta)\cos 2k\nu t,\\
			\sin(2\beta\sin\nu t)=&2\sum_{k=1}^{\infty}J_{2k-1}(2\beta)\sin(2k-1)\nu t.
		\end{align}
	\end{subequations}
	Under the conditions $\mu=2\nu$ and $\mu\gg A/2, g_0/2$, the Hamiltonian Eq.~\eqref{eqHamU2} reduces to the AQRM Eq.~\eqref{eqAQRM}, where the parameters are defined as $\Omega=AJ_2(2\beta)+(\omega_q-\omega_0)J_0(2\beta)$, $\omega=\omega_r-\omega_0$, $g_1=g_0(1+J_0(2\beta))/2$, $g_2=g_0(1-J_0(2\beta))/2$ and $\varepsilon=2V_0$. The anisotropy ratio is given by
	\begin{equation}
	    \lambda=\frac{1-J_0(2\beta)}{1+J_0(2\beta)},
	\end{equation}
	which has a tunable range from $0$ to $2.347$, as plotted in Fig.~\ref{tunable_range}.

	\subsubsection{Duality mapping}
	
	A Pauli $\hat{\sigma}_x$ transformation maps the rotating and counterrotating channels onto each other:
	\begin{equation}
	    \hat{\sigma}_x(\hat{a}^\dagger\hat{\sigma}_-+\hat{a}\hat{\sigma}_+)\hat{\sigma}_x=\hat{a}^\dagger\hat{\sigma}_++\hat{a}\hat{\sigma}_-,
	\end{equation}
	This follows from $\hat{\sigma}_x\hat{\sigma}_\pm\hat{\sigma}_x=\hat{\sigma}_\mp$ and $\hat{\sigma}_x\hat{\sigma}_z\hat{\sigma}_x=-\hat{\sigma}_z$, and provides an exact mapping from the $\lambda>1$ sector to the $0<\lambda<1$ sector of the AQRM through
    \begin{equation} \hat{\sigma}_x\hat{H}(\omega, \Omega, g_1,g_2,\varepsilon)\hat{\sigma}_x =\hat{H}(\omega, -\Omega, g_2,g_1,\varepsilon). 
    \end{equation}
    Experimentally, this duality allows a target AQRM with anisotropy $\lambda$ to be realized by implementing its dual Hamiltonian at $1/\lambda$ with opposite qubit frequency, together with single-qubit $X$ gates before and after the evolution.
	 
	\subsection{Dynamics under anisotropic interactions}
	Here we discuss the dynamics generated by a purely anisotropic interaction,
	\begin{equation}
	    \hat{H}_{\textrm{int}}=g_1(\hat{a}^{\dagger}\hat{\sigma}_-+\hat{a}\hat{\sigma}_+)+g_2 (\hat{a}^{\dagger}\hat{\sigma}_++\hat{a}\hat{\sigma}_-),
	\end{equation}
	for the initial state $\ket{\psi_0}=\ket{0, e}$. In general, this evolution does not admit a simple closed-form expression. We therefore consider three limiting cases:
	\begin{itemize}
	    \item When $g_1=g$ and $g_2=0$, the Hamiltonian exhibits a $U(1)$ symmetry associated with conservation of the total excitation number. In this case, the state at time $t$ is
	\begin{equation}
	    \ket{\psi}=\cos gt\ket{0, e}-\mathrm{i}\sin gt\ket{1, g},
	\end{equation}
    which clearly preserves the total excitation number of $N=1$.  
    \item For $g_1=g_2=g$, the state evolves into an entangled Schr\"odinger cat state:
    \begin{subequations}
        \begin{align}
            \ket{\psi}=&\mathrm{e}^{-\mathrm{i}gt\hat{\sigma}_x(\hat{a}+\hat{a}^{\dagger})}\frac{1}{\sqrt{2}}\ket{0,+}+\ket{0,-}\\
            =&\frac{1}{\sqrt{2}}(\ket{-\mathrm{i}gt}\otimes\ket{+}+\ket{\mathrm{i}gt}\otimes\ket{-})\\
            =&\frac{1}{2}[(\ket{-\mathrm{i}gt}+\ket{\mathrm{i}gt})\ket{e}+(\ket{-\mathrm{i}gt}-\ket{\mathrm{i}gt})\ket{g}]\\
            =&\frac{1}{2}(\mathcal{N}_+\ket{\mathcal{C}^+}\ket{e}+\mathcal{N}_-\ket{\mathcal{C}^-}\ket{g}),
        \end{align}
    \end{subequations}
    where $\ket{-\mathrm{i}gt}=\mathcal{D}(-\mathrm{i}gt)\ket{0}$ is a coherent state with displacement operator $\mathcal{D}(\alpha)=\mathrm{e}^{\alpha\hat{a}^{\dagger}-\alpha^*\hat{a}}$, and $\ket{\mathcal{C}^{\pm}}=(\ket{-\mathrm{i}gt}\pm\ket{\mathrm{i}gt})/\mathcal{N}_{\pm}$ are photonic cat states with normalization $\mathcal{N}_{\pm}=\sqrt{2(1\pm \mathrm{e}^{-2g^2t^2})}$. The total excitation number varies in time as
    \begin{equation}
        \langle\hat{N}\rangle=g^2t^2\coth(2g^2t^2)+\frac{1}{2},
    \end{equation}
    and the resonator photon number is $\langle\hat{a}^{\dagger}\hat{a}\rangle=g^2t^2\coth(2g^2t^2)$, exhibiting asymptotic growth proportional to $t^2$.
    \item When $g_1=0$ and $g_2=g$, since $\hat{a}^{\dagger}\hat{\sigma}_+\ket{0,e}=\hat{a}\hat{\sigma}_-\ket{0,e}=0$, the state remains in $\ket{0, e}$ with a constant total excitation number of $1$.
	\end{itemize}
	
	For $0<\lambda<1$ and $\lambda>1$, the total excitation number $N$ smoothly interpolates between a constant value and quadratic growth in time. For a fixed evolution time $t$, $N$ reaches its peak as a function of $\lambda$ when $g_1+g_2=J$ is held constant. Rewriting the Hamiltonian as
	\begin{equation}
	    \hat{H}_{\lambda}=\frac{J}{2}(\hat{H}_\textrm{r}+\hat{H}_{\textrm{cr}})+\frac{J(1-\lambda)}{2(1+\lambda)}(\hat{H}_{\textrm{r}}-\hat{H}_{\textrm{cr}}).
	\end{equation} 
	we define
	\begin{equation}
	    F(\lambda)=\bra{0, e}\mathrm{e}^{\mathrm{i}\hat{H}_{\lambda}t}\hat{N}\mathrm{e}^{-\mathrm{i}\hat{H}_{\lambda}t}\ket{0,e},
	\end{equation}
	and the corresponding derivative at $\lambda=1$ is given by
	\begin{equation}\label{eq_deriv}
	   F'(1)
	   =-\mathrm{Re}\left[\int_0^1\!\mathrm{d}\alpha\;\bra{0,e}\mathrm{e}^{\mathrm{i}\alpha\frac{Jt}{2}(\hat{H}_{\mathrm{r}}+\hat{H}_{\textrm{cr}})}(\hat{H}_{\textrm{r}}-\hat{H}_{\textrm{cr}})\mathrm{e}^{\mathrm{i}(1-\alpha)\frac{Jt}{2}(\hat{H}_{\mathrm{r}}+\hat{H}_{\textrm{cr}})}\hat{N}\mathrm{e}^{-\mathrm{i}\frac{Jt}{2}(\hat{H}_{\textrm{r}}+\hat{H}_{\mathrm{cr}})}\ket{0,e}\right],
	\end{equation}
	where we have used Wilcox's formula
	\begin{equation}
	    \frac{\mathrm{d}}{\mathrm{d}t}\mathrm{e}^{\hat{X}(t)}=\int_0^1\!\mathrm{d}\alpha\;\mathrm{e}^{\alpha \hat{X}(t)}\frac{\mathrm{d}}{\mathrm{d}t}\hat{X}(t)\mathrm{e}^{(1-\alpha)\hat{X}(t)}.
	\end{equation}
	The left vector can be calculated in the basis $\{\ket{+}, \ket{-}\}$ as
	\begin{subequations}
	\begin{align}
    	\bra{\Phi}=&\bra{0,e}\mathrm{e}^{\mathrm{i}\alpha\frac{Jt}{2}(\hat{H}_{\mathrm{r}}+\hat{H}_{\textrm{cr}})}(\hat{H}_{\textrm{r}}-\hat{H}_{\textrm{cr}})\mathrm{e}^{-\mathrm{i}\alpha\frac{Jt}{2}(\hat{H}_{\mathrm{r}}+\hat{H}_{\textrm{cr}})}\\
    	=&\bra{0,e}\mathrm{e}^{\mathrm{i}\alpha\frac{Jt}{2}(\hat{H}_{\mathrm{r}}+\hat{H}_{\textrm{cr}})}\mathrm{i}(\hat{a}-\hat{a}^{\dagger})\hat{\sigma}_y\mathrm{e}^{-\mathrm{i}\alpha\frac{Jt}{2}(\hat{H}_{\mathrm{r}}+\hat{H}_{\textrm{cr}})}\\
    	=&\frac{1}{\sqrt{2}}\begin{bmatrix}
    	    \bra{0} & \bra{0}
    	\end{bmatrix}
    	\begin{bmatrix}
    	&-\mathcal{D}(\mathrm{i}\alpha\frac{Jt}{2})(\hat{a}-\hat{a}^{\dagger})\mathcal{D}(\mathrm{i}\alpha\frac{Jt}{2})\\
    	\mathcal{D}(-\mathrm{i}\alpha\frac{Jt}{2})(\hat{a}-\hat{a}^{\dagger})\mathcal{D}(-\mathrm{i}\alpha\frac{Jt}{2})&
    	\end{bmatrix},\\
	\end{align}
	\end{subequations}
	while the right vector becomes
	\begin{subequations}
	\begin{align}
	    \ket{\Psi}=&\mathrm{e}^{\mathrm{i}\frac{Jt}{2}(\hat{H}_{\mathrm{r}}+\hat{H}_{\textrm{cr}})}\hat{N}\mathrm{e}^{-\mathrm{i}\frac{Jt}{2}(\hat{H}_{\textrm{r}}+\hat{H}_{\mathrm{cr}})}\ket{0,e}\\
	    =&\begin{bmatrix}
	    \mathcal{D}(\mathrm{i}\frac{Jt}{2})(\hat{a}^{\dagger}\hat{a}+\frac{1}{2})\mathcal{D}(-\mathrm{i}\frac{Jt}{2})& \frac{1}{2}\mathcal{D}(\mathrm{i}Jt)\\
	    \frac{1}{2}\mathcal{D}(-\mathrm{i}Jt) &\mathcal{D}(-\mathrm{i}\frac{Jt}{2})(\hat{a}^{\dagger}\hat{a}+\frac{1}{2})\mathcal{D}(\mathrm{i}\frac{Jt}{2})
	    \end{bmatrix}\frac{1}{\sqrt{2}}
	    \begin{bmatrix}
	        \ket{0} \\ \ket{0}
	    \end{bmatrix}.
	\end{align}
	\end{subequations}
	The integrand in Eq.~\eqref{eq_deriv} then evaluates to
	\begin{equation}
	    G(\alpha)=\bra{\Phi}\Psi\rangle=\frac{\mathrm{i}}{2}Jt[\mathrm{e}^{-\frac{1}{2}|(1-\alpha)Jt|^2}-\mathrm{e}^{-\frac{1}{2}|\alpha Jt|^2}],
	\end{equation}
	which is purely imaginary and satisfies $G(\alpha)+G(1-\alpha)=0$, leading to $F'(1)=0$. Thus, the total excitation number reaches an extremum at $\lambda=1$; the numerical dynamics show that this extremum is the peak.

	\subsection{Collapses and revivals in the USC/DSC regimes}
    Collapses and revivals of the initial population were first predicted in the JCM regime for a coherent initial state. This phenomenon disappears with increasing $g/\omega$ but reemerges in the ultrastrong coupling (USC) and deep-strong coupling (DSC) regimes for a Fock initial state. Here we examine the corresponding behavior in the AQRM using a procedure analogous to Ref.~\cite{Casanova2010}.

    In the unbiased AQRM the $\mathbb{Z}_2$ parity is conserved, so the dynamics is confined to two disconnected parity chains
    \begin{subequations}
    \begin{align}
    |0, g\rangle \leftrightarrow |1, e\rangle \leftrightarrow |2, g\rangle \leftrightarrow |3, e\rangle \cdots 
    \qquad &(p=+1),\\
    |0, e\rangle \leftrightarrow |1, g\rangle \leftrightarrow |2, e\rangle \leftrightarrow |3, g\rangle \cdots 
    \qquad &(p=-1),
    \end{align}
    \end{subequations}
    where neighboring links are generated by rotating and counter-rotating processes with generally different amplitudes. Introducing the parity basis $\ket{p,n_b}$ and the composite operator $\hat b=\hat a\hat{\sigma}_x$, the AQRM can be written as
    \begin{subequations}
    \begin{align}
    \hat{H}
    &=\omega\hat{a}^{\dagger}\hat{a}+\frac{\Omega}{2}\hat{\sigma}_z
    +g_x(\hat{a}+\hat{a}^{\dagger})\hat{\sigma}_x
    +\mathrm{i}g_y(\hat{a}-\hat{a}^{\dagger})\hat{\sigma}_y, \\
    &=\omega\hat{b}^{\dagger}\hat{b}
    -\frac{\Omega}{2}(-1)^{\hat{b}^{\dagger}\hat{b}}\Pi
    +g_x(\hat{b}+\hat{b}^{\dagger})
    +g_y(-1)^{\hat{b}^{\dagger}\hat{b}}\Pi(\hat{b}^{\dagger}-\hat{b}),
    \end{align}
    \end{subequations}
    which makes explicit that anisotropy introduces an additional orthogonal coupling channel (the $g_y$ term) within each parity sector.
    
    For a given initial state $\ket{\psi(0)}$, the revival probability can be expressed as
    \begin{equation}
    P_r(t)=\left|\sum_l \left|\braket{\psi(0)}{\phi_l}\right|^2 e^{-\mathrm{i}E_l t}\right|^2,
    \end{equation}
    where $\{\ket{\phi_l}\}$ and $\{E_l\}$ are eigenstates and eigenenergies of $\hat H$ in the relevant parity sector.
    
    At the isotropic point and in the degenerate case $\Omega=0$, the Hamiltonian reduces to a displaced oscillator,
    \begin{equation}
    \hat{H}_{\rm iso}=\omega \hat{b}^{\dagger}\hat{b}+g_x(\hat{b}+\hat{b}^\dagger)
    =\omega\left(\hat{b}^\dagger+\frac{g_x}{\omega}\right)\left(\hat{b}+\frac{g_x}{\omega}\right)-\frac{g_x^2}{\omega},
    \end{equation}
    yielding an exactly equally spaced spectrum
    \begin{equation}
    E_n=n\omega-\frac{g_x^2}{\omega}, \qquad n=0,1,2,\dots,
    \end{equation}
    and hence exact periodic revival with $T=2\pi/\omega$.
    For generic anisotropy ($g_y\neq 0$), this displaced-oscillator reduction no longer applies: the orthogonal channel reshapes the dressed spacings within each parity sector, so the spectrum is no longer an elementary equally spaced ladder. Nevertheless, the AQRM remains exactly solvable in the standard sense: in each parity sector one can construct a transcendental $G$-function whose zeros determine the spectrum implicitly~\cite{Xie2014},
    \begin{equation}
    G_{\pm}(E)=0,
    \end{equation}
    with $\pm$ labeling the two parity sectors (together with possible exceptional solutions associated with level crossings). Like the nondegenerate case of the QRM, when $\lambda\sim 1$, energy levels of AQRM deviate slightly from the equal-spaced distribution, leading to incomplete revivals of wave packets. 
	
	\subsection{Hidden symmetry and tunneling dynamics}
	
	\subsubsection{Spectra of AQRM when $\varepsilon=0$}
	
	We first discuss the spectrum of the AQRM when parity symmetry is conserved, i.e., $\varepsilon=0$. As in the standard QRM~\cite{Braak2011}, the spectrum contains regular and exceptional parts. The regular spectrum is defined by the zeros of a transcendental function and can be explicitly labeled by the two eigenvalues of the parity operator. The exceptional spectrum, by contrast, is characterized by doubly degenerate level crossings in the parameter space $(\omega, \Omega, g_1, g_2)$. These level crossings occur at energies
    \begin{equation}
	    E_n=n\omega-\frac{(1+\lambda)^2g_1^2}{2\omega},
	\end{equation}
	with integers $n=0, 1, 2, \cdots$. In particular, the case of $n=0$ corresponds to the crossing between the ground state and the first excited state. This specific degeneracy arises when the parameters satisfy the analytic condition:
	\begin{equation}
	    g_{1c}=\sqrt{\frac{\Omega\omega}{1-\lambda^2}}.
	\end{equation}
	Notably, such a crossing is forbidden in the isotropic QRM when $\lambda = 1$.
	
	\subsubsection{Exact solutions of AQRM}
	
	In general, the eigenvalues of the AQRM are determined by the zeros of the $G$-function, $G_{\varepsilon}(x) = 0$, where $E_n = x_n - \lambda g^2/\omega$. This function is constructed from infinite power series whose coefficients $K_n$ satisfy a three-term recurrence relation,
    \begin{equation}
    	a_n(x) K_{n+1} = b_n(x) K_n + c_n(x) K_{n-1}
    \end{equation}
    The complexity of this exact approach arises because the energy $E$ is embedded in the coefficients $a_n, b_n, c_n$. Solving for the spectrum therefore requires finding the roots of a highly nonlinear transcendental function, which obscures the intuitive mechanism behind the tunneling dynamics.
    
    For each root $x_n$, the corresponding exact eigenstate $\ket{\Psi_n}$ is an entangled state of the qubit and the displaced cavity field. In the original representation, it can be expressed as
    \begin{equation}
    	\ket{\Psi_n} = U(\lambda, \theta) \begin{pmatrix} \phi_1(x_n) \\ \phi_2(x_n) \end{pmatrix} = U(\lambda, \theta) \sum_{m=0}^{\infty} \begin{pmatrix} L_m^+(x_n) \\ K_m^+(x_n) \end{pmatrix} \ket{m}_B
    \end{equation}
    where $\ket{m}_B$ represents the basis states in Bargmann space and $U(\lambda, \theta)$ is defined as
     \begin{equation}
    	U(\lambda, \theta) = \frac{1}{\sqrt{1+\lambda}} \begin{pmatrix} 1 & -\sqrt{\lambda} \\ \sqrt{\lambda} & 1 \end{pmatrix}
    \end{equation}
    (Note: In this derivation, we set the phase factor $\theta=0$, consistent with the preceding sections). The component functions $\phi_1(z)$ and $\phi_2(z)$ are defined by infinite power series:
    \begin{align}
    	\phi_1(z) &= \exp\left( -\frac{\sqrt{\lambda}g}{\omega_c}z \right) \sum_{m=0}^{\infty} L_m^+ \left( z + \frac{\sqrt{\lambda}g}{\omega_c} \right)^m \\
    	\phi_2(z) &= \exp\left( -\frac{\sqrt{\lambda}g}{\omega_c}z \right) \sum_{m=0}^{\infty} K_m^+ \left( z + \frac{\sqrt{\lambda}g}{\omega_c} \right)^m
    \end{align}
    The coefficients $L_m^+$ and $K_m^+$ are determined by
    \begin{equation}
    	L_m^+ = \frac{f K_m^+ - (1-\lambda)g(m+1) K_{m+1}^+}{m\omega_c - \mathcal{E} + c}
    \end{equation}
    where $f = d + (1-\lambda)\sqrt{\lambda}g^2/\omega_c$. The coefficients $K_m^+$ satisfy
    \begin{equation}
    	a_n(\mathcal{E}) K_{n+1}^+ = b_n(\mathcal{E}) K_n^+ + c_n(\mathcal{E}) K_{n-1}^+
    \end{equation}
    with $K_{-1}^+ = 0, K_0^+ = 1$. The recurrence coefficients are
    \begin{align}
    	a_n(\mathcal{E}) &= 2\sqrt{\lambda} - \frac{(1-\lambda) f }{n\omega_c - \mathcal{E} + c} (n+1)g \\
    	b_n(\mathcal{E}) &= \frac{4\lambda g^2}{\omega_c} + n\omega_c - \mathcal{E} - c - \frac{f^* f}{n\omega_c - \mathcal{E} + c} - \frac{(1-\lambda)^2 g^2 n}{(n-1)\omega_c - \mathcal{E} + c} \\
    	c_n(\mathcal{E}) &= -2\sqrt{\lambda}g + \frac{(1-\lambda) g f }{(n-1)\omega_c - \mathcal{E} + c}
    \end{align}
    	
    The complexity of this exact approach is two-fold:
    \begin{enumerate}
    	\item \textit{Mathematical Intractability:} The energy $E$ is not an explicit variable but is buried within the coefficients of an infinite series. Finding the roots $x_n$ requires high-precision numerical searches for the zeros of $G_{\varepsilon}(x)$.
    	\item \textit{Physical Obscurity:} Since $\ket{\Psi_n}$ involves an infinite summation over displaced Fock states, the intuitive mechanism of tunneling (e.g., the transition between two specific potential wells) is hidden.
    \end{enumerate}
    This motivates the effective-Hamiltonian treatment in the following section, where a controlled approximation truncates this complexity and exposes the tunneling dynamics more directly.

    \subsubsection{Selective tunneling}
	
	When introducing a bias term, the isotropic QRM is promoted to the biased QRM. The bias explicitly breaks the $\mathbb{Z}_2$ parity symmetry, so the familiar parity-resolved decomposition into disconnected chains no longer applies. Remarkably, as emphasized by \cite{Mangazeev2021}, this does \emph{not} imply a featureless spectrum or dynamics: at special bias points $\varepsilon/\omega\in\mathbb{Z}$, the AQRM exhibits a hidden symmetry that protects level crossings and imposes additional selection rules. A direct dynamical consequence is the emergence of \emph{selective tunneling}: preparing a wave packet localized on one displaced-oscillator branch, the time evolution can display strongly enhanced or strongly suppressed tunneling depending on whether the bias is tuned to (or away from) the hidden-symmetry condition. Operationally, this can be quantified by the branch (or target-state) overlap
    \begin{equation}
    P_{\rm tun}(t)=\bigl|\langle \psi_{\rm target}|e^{-iH_{\rm AQRM}t}|\psi(0)\rangle\bigr|^2,
    \label{eq:Ptun}
    \end{equation}
    which develops a pronounced tunneling oscillation in the generic biased case, but becomes qualitatively reorganized at the hidden-symmetry points in accordance with the protected crossings and the associated selection rule~\cite{Li2021}. In this way, the biased model provides a sharp contrast to the unbiased QRM: instead of being organized by an explicit parity conservation, the spectrum and the tunneling dynamics are reorganized by a discrete set of bias-tuned hidden-symmetry conditions.
    
    While the biased isotropic Rabi model reveals hidden symmetry at discrete bias values, a richer structure emerges when the coupling becomes anisotropic, i.e., when the rotating and counterrotating terms have distinct strengths $g_{\rm r} \neq g_{\rm cr}$. As discussed above for the unbiased AQRM, coherent tunneling proceeds through a squeezed phase space with a deflected spin-polarization axis. When a bias is added to the anisotropic model, the hidden-symmetry condition becomes a function of both $\varepsilon$ and $\lambda$, leading to a two-dimensional discrete set of protected level crossings. Consequently, the tunneling dynamics displays a richer form of selectivity: by tuning either the bias or the anisotropy ratio, one can switch among enhanced, suppressed, and nearly blocked tunneling. This interplay between bias-tuned and anisotropy-tuned hidden symmetries opens a route to controlling quantum tunneling in circuit-QED and trapped-ion systems, where both anisotropy and bias are experimentally accessible.
    
    \subsubsection{Approximate tunneling states}

    To handle the bias term $\varepsilon \hat{\sigma}_x$ and the anisotropic coupling, we apply a unitary transformation $U_1$ to the qubit subspace~\cite{Xie2014}:
    \begin{equation}
    	U_1 = \frac{1}{\sqrt{1+\lambda}} \begin{pmatrix} 1 & -\sqrt{\lambda} \\ \sqrt{\lambda} & 1 \end{pmatrix}
    \end{equation}
    The transformed Hamiltonian $H_2 = U_1^\dagger H_1 U_1$ takes the form
    \begin{equation}
    	H_2 = \begin{pmatrix} \omega a^\dagger a + \sqrt{\lambda}g_1(a + a^\dagger) + c & (1-\lambda)g_1a - d \\ (1-\lambda)g_1a^\dagger - d & \omega a^\dagger a - \sqrt{\lambda}g_1(a + a^\dagger) - c \end{pmatrix}
    \end{equation}
    where the coefficients are
    \begin{equation}
    	c = \frac{1-\lambda}{1+\lambda} \frac{\Delta}{2} + \frac{\sqrt{\lambda}}{1+\lambda} \varepsilon, \quad d = \frac{\sqrt{\lambda}}{1+\lambda} \Delta - \frac{1-\lambda}{1+\lambda} \frac{\varepsilon}{2}
    \end{equation}
    	
    To eliminate the longitudinal coupling terms $\sqrt{\lambda}g_1(a+a^\dagger) \hat{\sigma}_z$, we introduce a conditional displacement operator,
    \begin{equation}
    	U_2 = \exp\left[ -\frac{\sqrt{\lambda}g_1}{\omega} \hat{\sigma}_z (a^\dagger - a) \right]
    \end{equation}
    This operator maps the system into the \textit{displaced oscillator basis} $\ket{n_\pm, \pm}$, where the cavity field equilibrium depends on the qubit state.

    The final Hamiltonian $H_3 = U_2^\dagger H_2 U_2$ is
    \begin{equation}
    	H_3 = \omega a^\dagger a - \frac{\lambda g_1^2}{\omega} + c \hat{\sigma}_z + \hat{\sigma}_+ (\tilde{g}a - x) e^{k(a^\dagger - a)} + \hat{\sigma}_- (\tilde{g}a^\dagger + y) e^{-k(a^\dagger - a)}
    	\label{eq:H3_exact}
    \end{equation}
    with $\tilde{g} = (1-\lambda)g_1$, $k = 2\sqrt{\lambda}g_1/\omega$, and parameters $x=(1-\lambda)\sqrt{\lambda}g_1^2/\omega+d, y=(1-\lambda)\sqrt{\lambda}g_1^2/\omega-d$. The exponential terms $e^{\pm k(a^\dagger - a)}$ introduce nonlinearities that couple all photon-number states.
    
    In the limit $k \ll 1$, we perform a first-order expansion of the exponential displacement operators, $e^{\pm k(a^\dagger - a)} \approx 1+k(a^\dagger - a)$. The system reduces to an analytical effective Jaynes-Cummings model with bias terms:
    \begin{equation}
    	H_3 \approx \omega a^\dagger a + c \hat{\sigma}_z +  (\tilde{g}+kx)(a \hat{\sigma}_+ +a^\dagger \hat{\sigma}_-)+(y+k\tilde{g}a^\dagger a)\hat{\sigma}_x. \label{eqre}
    \end{equation}
    This approximation remains valid for small anisotropy and weak to moderate coupling, providing a clear intuitive picture of the underlying dynamics.
    
    In the displaced oscillator basis, the $\hat{\sigma}_\pm$ operators act as tunneling operators between the two spatially separated potential wells. The constant term $(y+k\tilde{g}a^\dagger a)\hat{\sigma}_x$ drives direct tunneling that preserves photon number, whereas the interaction term $(\tilde{g}+kx)(a \hat{\sigma}_+ +a^\dagger \hat{\sigma}_-)$ enables photon-assisted tunneling that changes photon number. The coherent oscillations observed in our simulations directly reflect these tunneling processes at the hidden-symmetry points.
    
    Tunneling in the biased AQRM is selective. Unlike ordinary QRM oscillations, it occurs predominantly when the bias $\varepsilon$ satisfies the resonance condition. In the reduced model in Eq.~(\ref{eqre}), this condition is
    \begin{equation}
    	\varepsilon = \frac{1+\lambda}{2\sqrt{\lambda}} n \omega - \frac{1-\lambda}{2\sqrt{\lambda}} \Delta, \quad n = 0, 1, 2, \dots
    \end{equation}
    At these values, the system exhibits hidden symmetry, leading to level crossings (Juddian points) in the energy spectrum.
    
    Under the resonance condition, the tunneling occurs between the correlated states:
    \begin{itemize}
    	\item For $n=0$ ($c=0$): $\ket{0_+, +} \leftrightarrow \ket{0_-, -}$,
    	\item For $n=1$ ($c=0.5\omega$): $\ket{0_+, +} \leftrightarrow \ket{1_-, -}$,
    \end{itemize}
    where
    \begin{eqnarray}
    	\ket{0_+, +} &=& \ket{-\frac{\sqrt{\lambda}g_1}{\omega}} \otimes \left( \ket{e} + \sqrt{\lambda}\ket{g} \right)/\sqrt{1+\lambda}, \\
    	\ket{0_-, -} &=& \ket{\frac{\sqrt{\lambda}g_1}{\omega}} \otimes \left( -\sqrt{\lambda}\ket{e} + \ket{g} \right)/\sqrt{1+\lambda}, \\
    	\ket{1_-, -} &=& D\left( \frac{\sqrt{\lambda}g_1}{\omega} \right) \ket{1} \otimes \left( -\sqrt{\lambda}\ket{e} + \ket{g} \right)/\sqrt{1+\lambda}.
    \end{eqnarray}
    The corresponding quasi-eigenenergies are split by the tunneling matrix element, which determines the frequency of the probability oscillations observed in numerical simulations.
    
    Because the analytical derivation above is primarily controlled in the weak-coupling regime ($g \le 0.2\omega$), the parameters obtained from the simplified model cannot be applied directly to the experimental strong-coupling setting, where clear tunneling dynamics are easier to resolve. We therefore use a numerical approach to identify optimal parameters for realizing tunneling dynamics in the strong-coupling regime ($g \sim \omega$).
    
    We assume that the low-energy tunneling processes are predominantly governed by a few-level subspace spanned by the displaced oscillator states $\ket{0_+, +}$, $\ket{0_-, -}$, and $\ket{1_-, -}$. We then simulate the tunneling dynamics under strong coupling. First, we plot the first ten energy levels as a function of the normalized bias parameter:
    \begin{equation}
    	\varepsilon = k\frac{2\sqrt{\lambda}}{1+\lambda}\omega,
    \end{equation}
    with integer $k$.
    
\begin{figure}
    	\centering
    	\includegraphics[width=0.7\linewidth]{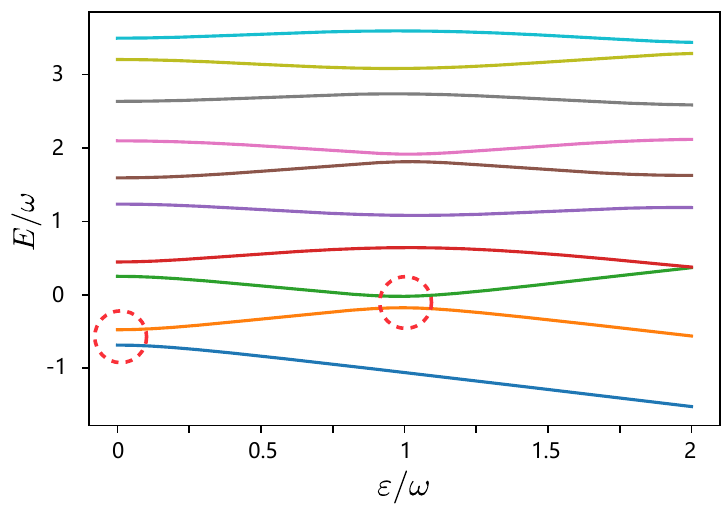}
    	\caption{Energy levels of the biased AQRM. The red dashed circles mark the tunneling processes. The parameters are $g_1=\omega$, $\Delta=0.1\omega$, and $\lambda=0.5$.}
    	\label{F2}
    \end{figure}

    The spectrum reveals pronounced avoided level crossings at $n=0$ and $n=1$, indicating the formation of a nearly closed dynamical subspace at these resonance points (see Fig.~\ref{F2}).
    
\begin{figure}
    	\centering
    	\includegraphics[width=0.99\linewidth]{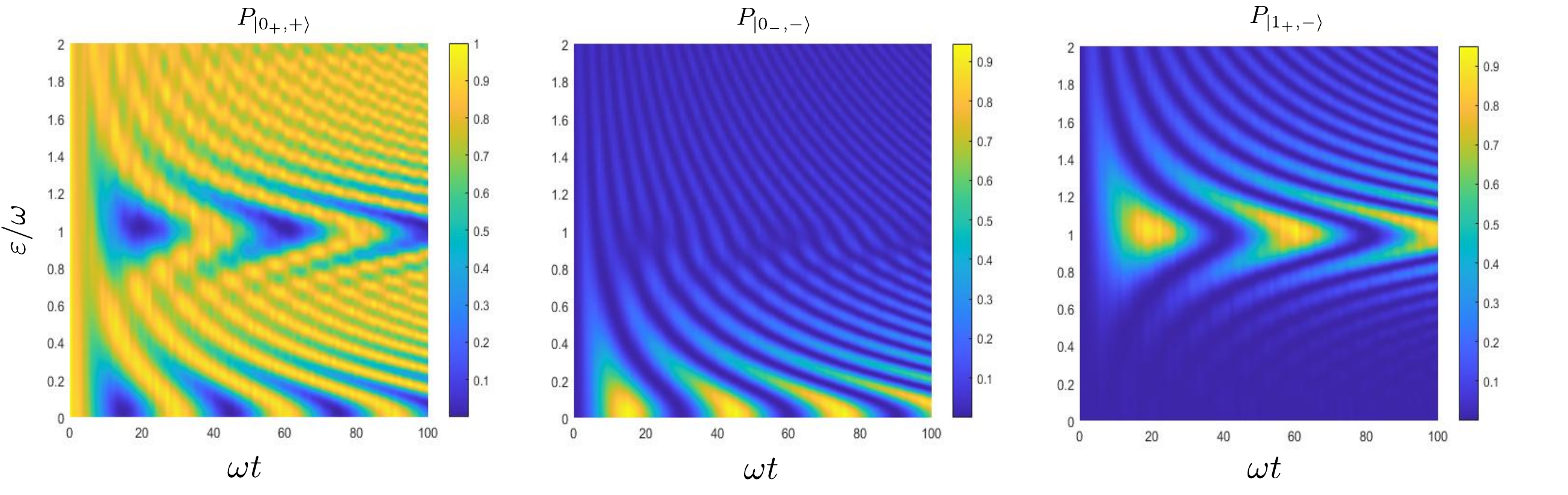}
    	\caption{Tunneling dynamics of the biased AQRM from the initial state $\ket{0_+, +}$. The population in the state (left) $\ket{0_+,+}$, (center) $\ket{0_-, -}$ and (right) $\ket{1_-, -}$. The parameters are $g=\omega$, $\Delta=0.1\omega$, and $\lambda=0.5$.}
    	\label{F3}
    \end{figure}

    To verify this assumption, we calculate the overlap (fidelity) between the three lowest eigenstates and the basis states $\ket{0_+, +}$, $\ket{0_-, -}$, and $\ket{1_-, -}$. The results confirm that these displaced states are the dominant components of the low-energy manifold during tunneling. Finally, we perform a parameter scan across different values of the normalized parameter $\bar{\varepsilon}$ to characterize the tunneling dynamics (see Fig.~\ref{F3}), providing a robust numerical guide for experimental observation in the strong coupling regime~\cite{Mangazeev2021}.
    
    	\begin{figure}[h]
	    \centering
	    \includegraphics[width=1.0\textwidth]{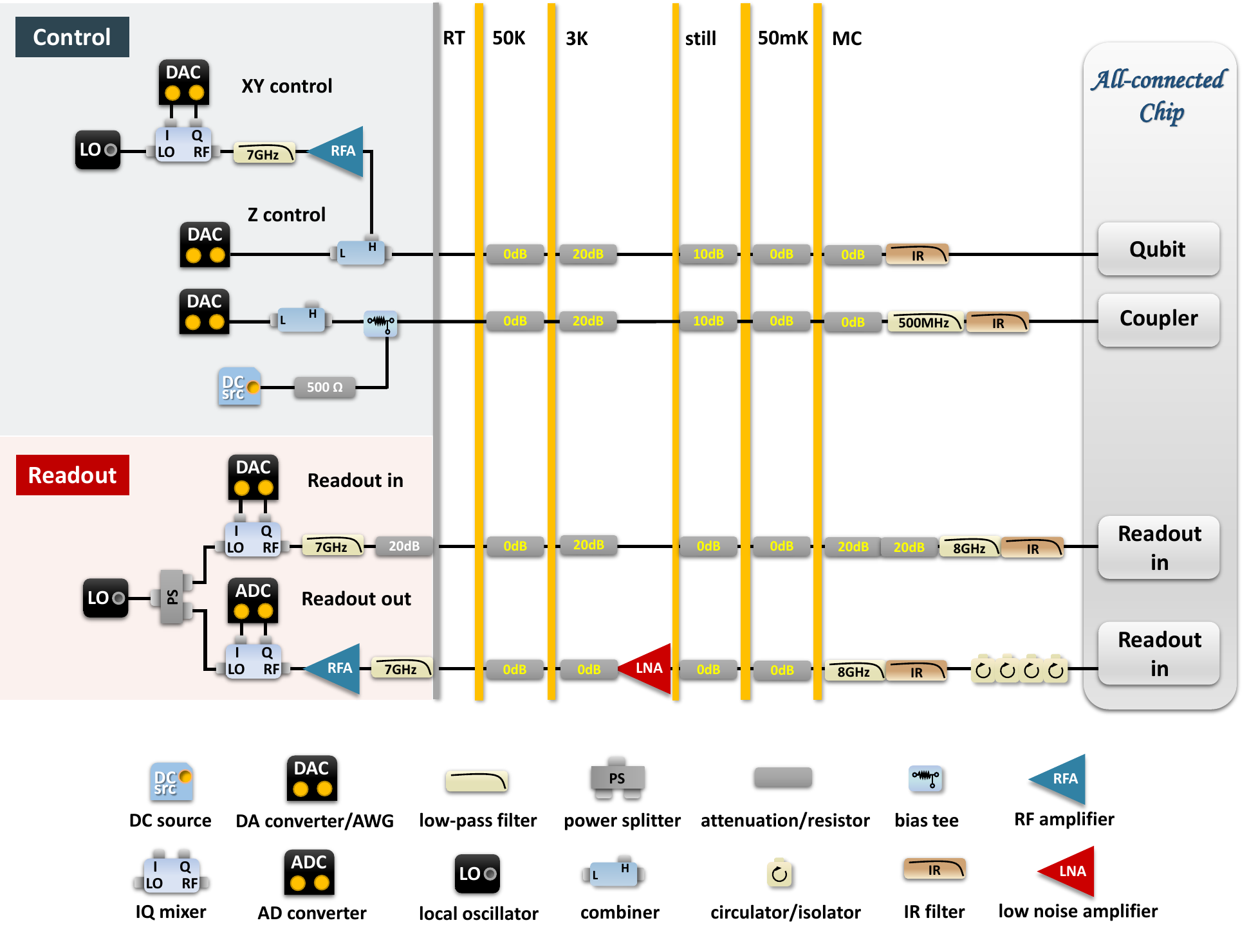}
	    \caption{\textbf{Schematic diagram of the experimental system and wiring information}}
	    \label{fig_wiring}
	\end{figure}
    
    \clearpage
    \newpage
	
	\section{Experimental Setup}
	\subsection{Wiring layout}
	
	Figure~\ref{fig_wiring} shows the wiring layout used for room-temperature and cryogenic operations. From top to bottom, the lines correspond to XY control, Z control (including fast bias and DC sources), and qubit readout (input and output) channels. Horizontally, the diagram shows the cooling stages from room temperature to 12~mK in a BlueFors XLD-1000 dilution refrigerator. Appropriate attenuation is implemented on the coaxial lines to suppress thermal noise~\cite{Krinner2019}.
	
	At room temperature, the high-frequency XY signals and low-frequency Z biases (fast Z) are combined and sent to the qubits for state excitation and frequency modulation, respectively. The XY signals are generated by IQ mixers, which mix local-oscillator (LO) signals from microwave sources with IQ waveforms from arbitrary waveform generators (AWGs). To deliver the strong transverse drive used in this experiment, an additional RF amplifier (RFA) is inserted into the XY control line at room temperature. On the coupler line, only the low-frequency Z bias is delivered. To extend the available bias range and reduce distortion of fast-Z pulses, the DC source is combined with the low-frequency Z bias through bias tees before reaching the coupler.
	
	Similarly, the readout signal is generated by mixing the AWG waveform with the LO signal through an IQ mixer. The signal collected from the readout output line is amplified by a low-noise amplifier (LNA) and a room-temperature RF amplifier (RFA). Before digitization, the RF signal is converted to an intermediate frequency through down-conversion and then sampled by analog-to-digital converters (ADCs).
	
	\subsection{Device performance}
	The device was fabricated using a flip-chip process and comprises 24 frequency-tunable transmon qubits. Among them, 22 experimental qubits are coupled to a central bus resonator $R$ through tunable couplers $C$, enabling in situ control of the effective qubit-resonator interaction. Two of these experimental qubits were designed with a large anharmonicity of approximately -0.37 GHz for anisotropic quantum Rabi model (AQRM) experiments, while the remaining 20 have an anharmonicity of approximately -0.2 GHz for other applications. In the AQRM measurements, one large-anharmonicity qubit is used as the system qubit $Q$, and an ancillary transmon qubit $Q_a$, directly coupled to $R$, is used for photon-number-resolved resonator readout. The relevant coherence and readout characteristics of $Q$, $Q_a$,and $R$ are summarized in Table S1; unless stated otherwise, these quantities are measured at the operating points used in the AQRM experiments.

    \begin{figure}
    	\centering
    	\includegraphics[width=0.7\linewidth]{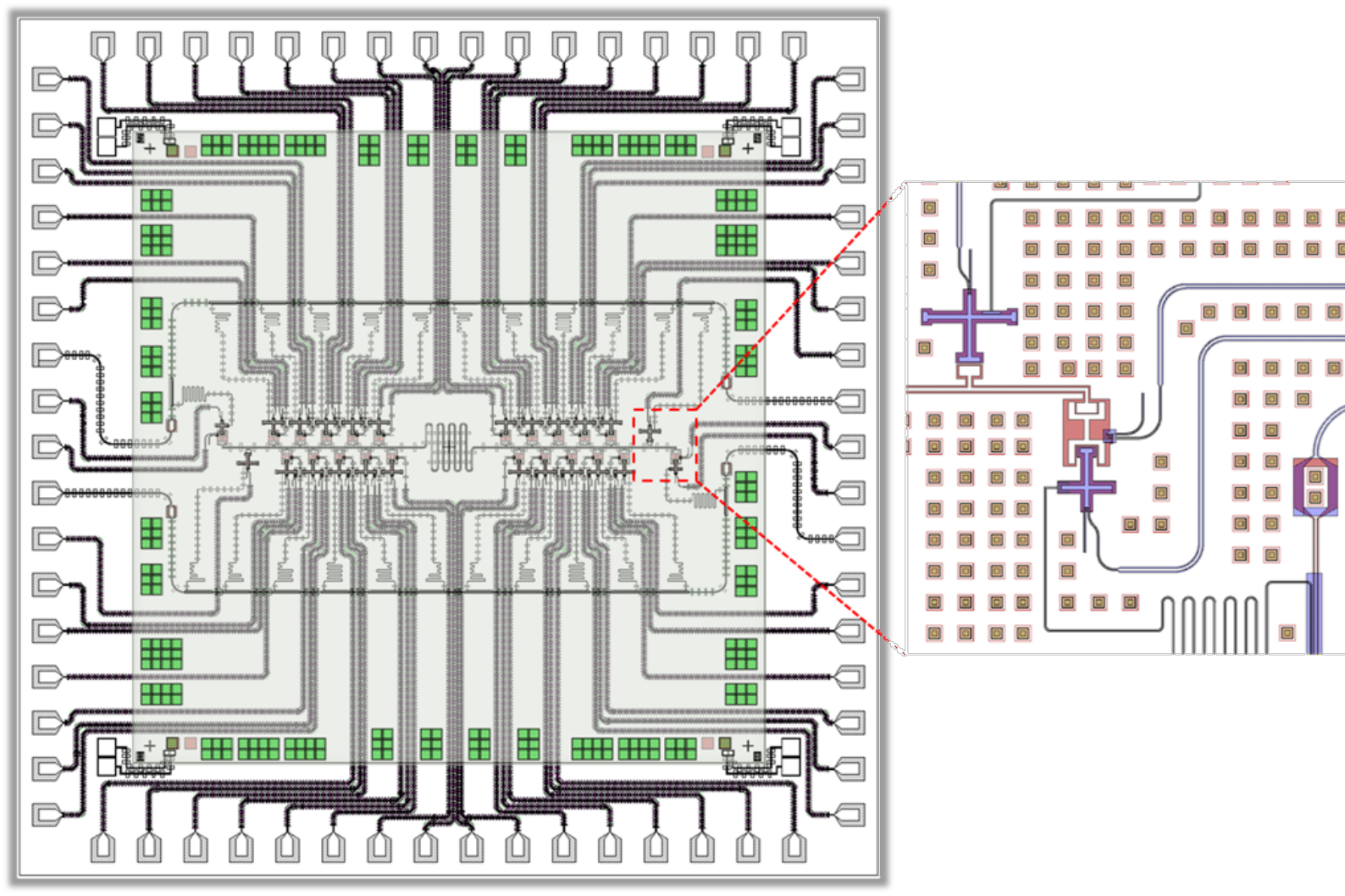}
    	\caption{\textbf{The micrograph of the superconducting quantum processor.}The right panel shows a magnified view of the circuit elements used for the anisotropic quantum Rabi model experiment, including the experimental system qubit $Q$, the tunable coupler $C$, the central bus resonator $R$, and the ancillary qubit $Q_a$ used for photon-number-resolved resonator readout.}
    	\label{Chip}
    \end{figure}

	\begin{table}[h]
	\centering
	\caption{\textbf{Device coherence and readout characteristics.} The symbols $Q$, $Q_a$, and $R$ denote the system qubit, ancilla qubit, and bus resonator, respectively. $T_1$, $T_2^*$, and $T_2^{\rm SE}$ are the energy relaxation time, Ramsey dephasing time, and echo dephasing time, respectively.}
	\label{tab:device_parameters}
	\begin{tabular}{lccc}
	\hline
	Parameter & $Q$ & $Q_a$ & $R$ \\
	\hline
	Operating frequency, $\omega_{10}/2\pi$ (GHz) & 4.420 & 4.454 & 3.966 \\
	Anharmonicity, $\alpha/2\pi$ (MHz) & -370 & -204 & -- \\
	Relaxation time, $T_1$ ($\mu$s) & 43.6 & 48.2 & 28.0 \\
	Ramsey dephasing time, $T_2^*$ ($\mu$s) & 10.3 & 13.5 & -- \\
	Echo dephasing time, $T_2^{\rm SE}$ ($\mu$s) & 20.9 & 23.1 & -- \\
	Readout fidelity of $\ket{g}$, $F_g$ (\%) & 96.83 & 97.60 & -- \\
	Readout fidelity of $\ket{e}$, $F_e$ (\%) & 93.80 & 93.73 & -- \\
	\hline
	\end{tabular}
	\end{table}
	
	For the experiments reported in the main text, the system qubit is operated in a regime where its anharmonicity suppresses leakage to higher transmon levels and supports the effective two-level description used in the AQRM Hamiltonian. The coherence times of the qubits and the resonator lifetime are longer than the relevant pulse sequences used for collapse-revival measurements, adiabatic eigenstate preparation, and tunneling oscillations, but still set the dominant decoherence limits in the experiment.
	
	
	\section{Experimental Control and Calibration}
	
	\subsection{Tunable coupler between qubit and resonator}
	
	\subsubsection{Circuit Hamiltonian}
	We consider a system in which the qubit and the resonator are coupled to a tunable coupler. The equivalent circuit is shown in Supplementary Fig.~\ref{equivalent circuit}.
    The resonator is modeled as an effective LC circuit.
    The circuit parameters satisfy the condition $C_{q},C_{c},C_{r}\gg C_{qc},C_{rc}\gg C_{qr}$,
    where $C_{qc}$ ($C_{rc}$) is the coupling capacitance between the qubit (resonator) and the coupler, and $C_{qr}$ is the direct capacitance between the qubit and the resonator.

	\begin{figure}[h]
	    \centering
	    \includegraphics[width=0.85\textwidth]{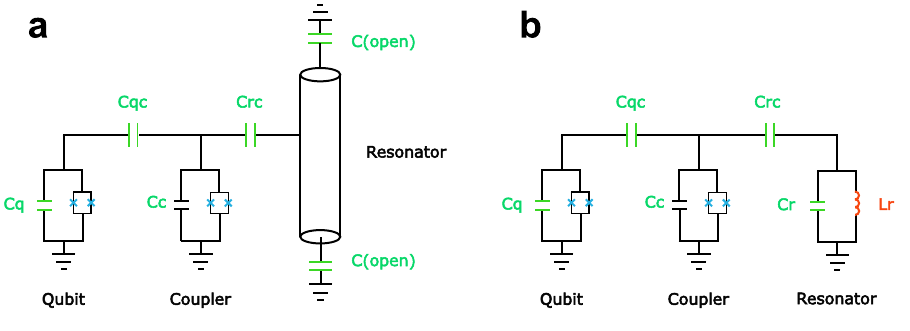}
	    \caption{ \textbf{a}, Equivalent circuit model for a qubit and a resonator coupled to a tunable coupler. \textbf{b}, Effective LC-circuit representation of the resonator. The effective capacitance and inductance depend on the respective coupling positions.}
	    \label{equivalent circuit}
	\end{figure}
	
    
    The system Lagrangian is composed of kinetic (\( T \)) and potential (\( U \)) energy terms, expressed using node fluxes \( \varphi_q \), \( \varphi_r \), and \( \varphi_c \). The kinetic energy is
    \begin{equation}
    \begin{aligned}
        T &=\frac{1}{2}\left[C_q \dot{\varphi_q}^2 + C_r \dot{\varphi_r}^2 + C_c \dot{\varphi_c}^2 \right] \\
        &+ \frac{1}{2} \left[C_{qc}\left(\dot{\varphi_q}- \dot{\varphi_c}\right)^2 +C_{rc}\left(\dot{\varphi_r}- \dot{\varphi_c}\right)^2 +C_{qr}\left(\dot{\varphi_q}- \dot{\varphi_r}\right)^2 \right].
    \end{aligned}
    \end{equation}

    The kinetic energy can be expressed in compact form as \( T = \frac{1}{2} \dot{\vec{\varphi}}^\top C \dot{\vec{\varphi}} \), where \( \vec{\varphi} = [\varphi_q, \varphi_r, \varphi_c]^\top \) and \( C \) is the \(3\times3\) capacitance matrix
    \begin{equation}
    \begin{aligned}
    C=\begin{bmatrix}
    C_q+C_{qr}+C_{qc} &  -C_{qr} & -C_{qc} \\
    -C_{qr} &  C_r+C_{qr}+C_{rc} & -C_{rc} \\
    -C_{qc} &  -C_{rc} & C_c+C_{qc}+C_{rc} \\
    \end{bmatrix}.
    \end{aligned}
    \end{equation}
    
    Under the condition \( C_q, C_c, C_r \gg C_{qc},C_{rc} \gg C_{qr} \), the inverse capacitance matrix is approximated as
    \begin{equation}
    \begin{aligned}
    C^{-1} \approx \begin{bmatrix}
    \frac{1}{C_q} &  \frac{ C_{qc} C_{rc}  + C_c C_{qr}}{C_{q} C_{r} C_{c}} & \frac{C_{qc}}{C_q C_c}\\
    \frac{C_{qc} C_{rc}  + C_c  C_{qr}}{C_{q} C_{r} C_{c}}  & \frac{1}{C_r} &  \frac{C_{rc}}{C_r C_c}\\
    \frac{C_{qc}}{C_q C_c} &  \frac{C_{rc}}{C_r C_c} & \frac{1}{C_c} \\
    \end{bmatrix}.
    \end{aligned}
    \end{equation}
    
    The conjugate charge vector is \( \vec{q} = C \dot{\vec{\varphi}} \), and the classical Hamiltonian is
    \begin{equation}
    H = \frac{1}{2} \vec{q}^\top C^{-1} \vec{q} + U. 
    \end{equation}
    Here, the potential energy $U$ includes the Josephson energies of the qubit and coupler and the inductive energy of the resonator:
     \begin{equation}
        \begin{aligned}
            U &= E_{J_q}\left[1-\cos\left(\frac{2\pi}{\Phi_0}\varphi_q\right)\right]+E_{J_c}\left[1-\cos\left(\frac{2\pi}{\Phi_0}\varphi_c\right)\right]+\frac{{\varphi_r}^2}{2L_r},
        \end{aligned}
    \end{equation}
    where \( E_{J_q} \) (\( E_{J_c} \)) is the Josephson energy of the qubit (coupler), and \( \Phi_0 = h / (2e) \) is the flux quantum. Quantizing the system, we promote \( \vec{q} \) and \( \vec{\varphi} \) to operators satisfying \( [\hat{\varphi}_\lambda, \hat{q}_{\lambda'}] = i \hbar \delta_{\lambda \lambda'} \). The quantized Hamiltonian is
    \begin{equation}
    \begin{aligned}
        \hat{H} = & 4E_{c_q}\left(\hat{n_q}\right)^2-E_{J_q}\cos\left(\frac{2\pi}{\Phi_0}\hat{\varphi_q}\right)
        +4E_{c_c}\left(\hat{n_c}\right)^2-E_{J_c}\cos\left(\frac{2\pi}{\Phi_0}\hat{\varphi_c}\right)    +4E_{c_r}\left(\hat{n_r}\right)^2+\frac{{\hat{\varphi_r}}^2}{2L_r}\\
        &+8\frac{C_{qc}}{\sqrt{C_q C_c}}\sqrt{E_{c_q}E_{c_c}}\left(\hat{n_q}\hat{n_c}\right)+8\frac{C_{rc}}{\sqrt{C_r C_c}}\sqrt{E_{c_r}E_{c_c}}\left(\hat{n_r}\hat{n_c}\right)\\
        &+8\left(\frac{  C_{qc} C_{rc} + C_c  C_{qr}}{ C_{c} \sqrt{C_{q} C_{r} }}\right)\sqrt{E_{c_q}E_{c_r}}\left(\hat{n_q}\hat{n_r}\right),\\
    \end{aligned}
    \end{equation}
    where \( \hat{n}_\lambda = \hat{q}_\lambda / (2e) \) is the Cooper-pair number operator, and \( E_{c_\lambda} = e^2 / (2 C_\lambda) \) is the charging energy for \( \lambda = q,c \).
    
    In the transmon regime (\( E_{J_\lambda} / E_{c_\lambda} \gg 1 \)), the qubits and resonator behave as weakly anharmonic oscillators. We approximate the system as coupled Duffing oscillators (\( \hbar = 1 \)):
    \begin{equation}
    \hat{H} = \hat{H}_q + \hat{H}_r + \hat{H}_c + \hat{H}_{qc} + \hat{H}_{rc} + \hat{H}_{qr}, 
    \end{equation}
    where the individual mode Hamiltonians are
    \begin{equation}
     \hat{H}_\lambda = \omega_\lambda \hat{b}_\lambda^\dagger \hat{b}_\lambda + \frac{\alpha_\lambda}{2} \hat{b}_\lambda^\dagger \hat{b}_\lambda^\dagger \hat{b}_\lambda \hat{b}_\lambda, \quad \lambda =q,r,c, 
     \end{equation}
    with anharmonicities
    \begin{equation}
     \alpha_\lambda = -E_{c_\lambda}, \quad \lambda = q,c. 
    \end{equation}
    The interaction terms are
    \begin{equation}
     \hat{H}_{jc} = g_{jc} \left( \hat{b}_j^\dagger \hat{b}_c + \hat{b}_j \hat{b}_c^\dagger - \hat{b}_j^\dagger \hat{b}_c^\dagger - \hat{b}_j \hat{b}_c \right), \quad j = q,r, 
    \end{equation}
    \begin{equation}
     \hat{H}_{qr} = g_{qr} \left( \hat{b}_q^\dagger \hat{b}_r + \hat{b}_q \hat{b}_r^\dagger - \hat{b}_q^\dagger \hat{b}_r^\dagger - \hat{b}_q \hat{b}_r \right), 
    \end{equation}
    with coupling strengths
    \begin{equation}
     g_{qc} = \frac{1}{2} \frac{C_{qc}}{\sqrt{C_q C_c}} \sqrt{\omega_q \omega_c}, 
    \end{equation}
    \begin{equation}
     g_{rc} = \frac{1}{2} \frac{ C_{rc}}{\sqrt{C_r C_c}} \sqrt{\omega_r \omega_c}, 
    \end{equation}
    \begin{equation}
     g_{qr} = \frac{1}{2} \frac{C_{qc} C_{rc} + C_{qr} C_c}{\sqrt{C_q C_r} C_c} \sqrt{\omega_q \omega_r} = \frac{1}{2} (1 + \eta) \frac{C_{qr}}{\sqrt{C_q C_r}} \sqrt{\omega_q \omega_r},
    \end{equation}
    where
    \begin{equation}
    \eta =  \frac{C_{qc} C_{rc}}{C_{qr} C_c}. 
    \end{equation}
    
    To eliminate the qubit-coupler and resonator-coupler interactions and obtain an effective qubit-resonator coupling, we apply the Schrieffer-Wolff transformation
    \begin{equation}
    \hat{U} = \exp \left\{ \sum_{j=q,r} \left[ \frac{g_{jc}}{\Delta_{jc}} \left( \hat{b}_j^\dagger \hat{b}_c - \hat{b}_j \hat{b}_c^\dagger \right) - \frac{g_{jc}}{\Sigma_{jc}} \left( \hat{b}_j^\dagger \hat{b}_c^\dagger - \hat{b}_j \hat{b}_c \right) \right] \right\},
    \end{equation}
    where
    \begin{equation}
    \Delta_{jc} = \omega_j - \omega_c, \Sigma_{jc} = \omega_j + \omega_c. 
    \end{equation}
    The transformed Hamiltonian is
    \begin{equation}
    \hat{\widetilde{H}} = \hat{U} \hat{H} \hat{U}^\dagger = \widetilde{\omega_q} \hat{b}_q^\dagger \hat{b}_q + \frac{\widetilde{\alpha_q}}{2} \hat{b}_q^\dagger \hat{b}_q^\dagger \hat{b}_q \hat{b}_q + \widetilde{\omega_r} \hat{b}_r^\dagger \hat{b}_r +  \widetilde{g}_{qr} \left( \hat{b}_q^\dagger \hat{b}_r + \hat{b}_q \hat{b}_r^\dagger \right),
    \end{equation}
    with
    \begin{equation}
    \widetilde{\omega_q} \approx \omega_q + g_{qc}^2 \left( \frac{1}{\Delta_{qc}} - \frac{1}{\Sigma_{qc}} \right), \widetilde{\alpha_q} \approx \alpha_q,
    \end{equation}
    \begin{equation}
    \widetilde{\omega_r} \approx \omega_r + g_{rc}^2 \left( \frac{1}{\Delta_{rc}} - \frac{1}{\Sigma_{rc}} \right),
    \end{equation}
    \begin{equation}
    \widetilde{g}_{qr} = \frac{g_{qc} g_{rc}}{2} \left( \frac{1}{\Delta_{qc}} + \frac{1}{\Delta_{rc}} - \frac{1}{\Sigma_{qc}} - \frac{1}{\Sigma_{rc}} \right) + g_{qr}.
    \end{equation}
    
    \subsubsection{Tunable coupling}
	The effective qubit-resonator coupling is controlled in situ by tuning the coupler frequency with a flux-bias pulse. Experimentally, we calibrate the mapping from the coupler control amplitude to the effective coupling strength $g_0$ before implementing the synthesized AQRM. The calibration uses the coherent exchange between the system qubit and the resonator near resonance: the qubit is prepared in $\ket{e}$ while the resonator is initialized close to vacuum, the coupler is biased to a selected operating point, and the resulting vacuum-Rabi oscillation is measured as a function of interaction time. Fitting the oscillation frequency yields the corresponding value of $g_0$.
	
\begin{figure}[h]
	    \centering
	    \includegraphics[width=0.95\textwidth]{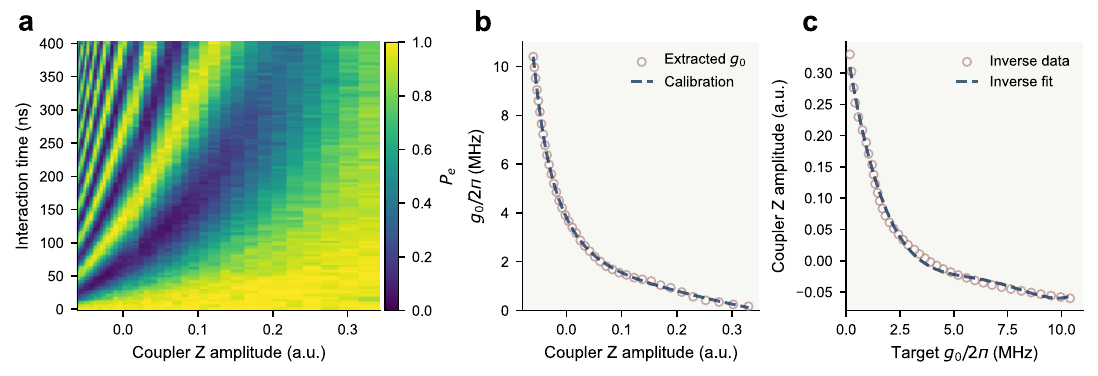}
	    \caption{\textbf{Calibration of the tunable qubit-resonator coupling.} \textbf{a}, Raw qubit-resonator exchange oscillations measured as a function of interaction time and coupler Z-pulse amplitude. \textbf{b}, Extracted effective coupling strength $g_0/2\pi$ and smooth calibration curve versus coupler Z-pulse amplitude. \textbf{c}, Inverse calibration used to convert a target coupling strength $g_0$ into the coupler Z-pulse amplitude for adiabatic ramps.}
	    \label{fig:tunable_coupling_calibration}
	\end{figure}

	The measured dependence $g_0$ versus the bias is then used to determine the operating points for the data shown in the main text. In the present experiment, the coupling $g_0/2\pi$ can be tuned from $-10$ to $0$~MHz, with the largest coupling used for the strongest-interaction data. For each target value of $g_0$, the coupler pulse amplitude and the associated frequency shifts of the qubit and resonator are calibrated and incorporated into the rotating-frame settings used for state preparation and tomography. The calibration workflow and inverse conversion are summarized in Fig.~\ref{fig:tunable_coupling_calibration}.
	
	 \subsection{Calibration of synthesized AQRM}
    \subsubsection{Calibration of control fields}
    The synthesized AQRM requires calibrated transverse qubit drives and longitudinal frequency modulation. The transverse terms are generated by microwave tones applied through the qubit XY line, including the carrier and the sideband tones used in the difference-frequency implementation. For each relevant tone, we calibrate the qubit response by measuring vacuum Rabi oscillations as a function of the applied microwave amplitude. The extracted response curve converts the programmed XY amplitude to the transverse drive strength used in the effective Hamiltonian.
    
\begin{figure}[h]
        \centering
        \includegraphics[width=0.95\textwidth]{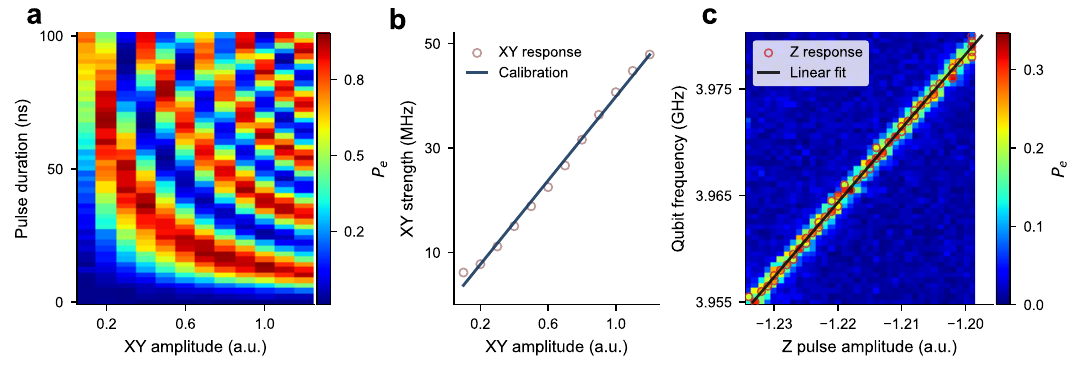}
        \caption{\textbf{Calibration of qubit control fields.} \textbf{a}, Raw transverse Rabi response measured as a function of XY amplitude and pulse duration. \textbf{b}, Extracted transverse response calibration, which converts the applied XY amplitude to the effective transverse drive strength. \textbf{c}, Longitudinal Z response obtained by measuring the qubit frequency shift versus Z-pulse amplitude. Over the small modulation range used in the experiment, the longitudinal response is well approximated by a linear calibration.}
        \label{fig:qubit_control_calibration}
    \end{figure}

    The longitudinal modulation is applied through the qubit Z line. Because the modulation amplitude used in the experiment is small compared with the full tunable range of the qubit, the local frequency response is well approximated by a linear dependence on the Z-pulse amplitude. The measured frequency shift is therefore fitted with a first-order calibration curve and used to set the longitudinal modulation amplitude. The corresponding transverse and longitudinal calibrations are shown in Fig.~\ref{fig:qubit_control_calibration}.
    
    \subsubsection{Displacement operators and resonator frequency}
\begin{figure}[h]
        \centering
        \includegraphics[width=0.85\textwidth]{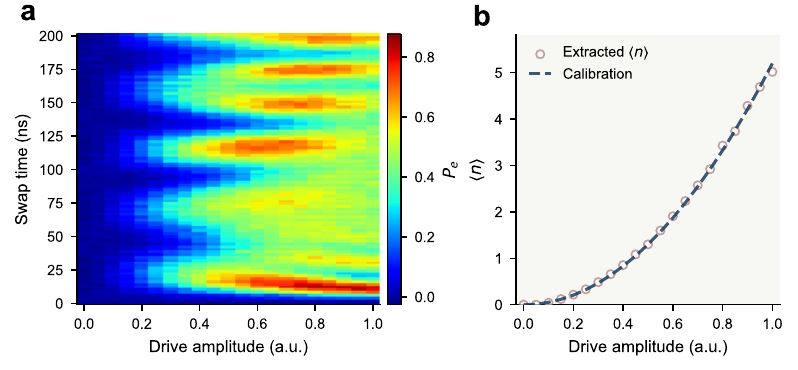}
        \caption{\textbf{Resonator drive calibration.} \textbf{a}, Photon-number-resolved readout response measured after resonator displacement pulses with different drive amplitudes. \textbf{b}, Extracted mean photon number $\langle n\rangle$ and calibration curve versus resonator drive amplitude. For a resonant displacement of the vacuum state, the displacement amplitude is calibrated using $|\alpha|=\sqrt{\langle n\rangle}$, giving the conversion between the applied resonator pulse amplitude and the phase-space displacement used in tomography and state preparation.}
        \label{fig:resonator_drive_calibration}
    \end{figure}

    Resonator displacements used for tomography, resonator-frequency calibration, and tunneling-state preparation are calibrated by measuring the photon number generated by resonant microwave pulses. For this calibration, the resonator pulse duration is fixed while its amplitude is varied. After each pulse, the photon-number distribution $P_n$ is extracted using the photon-number-resolved readout described above, and the mean photon number is calculated as
    \begin{equation}
        \langle n\rangle=\sum_n nP_n .
    \end{equation}
    For a resonant displacement of the vacuum state, the prepared state is a coherent state $D(\alpha)\ket{0}$ and satisfies
    \begin{equation}
        |\alpha|=\sqrt{\langle n\rangle}.
    \end{equation}
    This provides the calibration curve between the applied pulse amplitude and the displacement amplitude $|\alpha|$. The phase of the resonator drive sets the phase of $\alpha$, while the calibrated amplitude determines the displacement radius used in the tomography and state-preparation protocols. The raw readout response and the resulting displacement-amplitude calibration are shown in Fig.~\ref{fig:resonator_drive_calibration}.
    
\begin{figure}[h]
        \centering
        \includegraphics[width=0.9\textwidth]{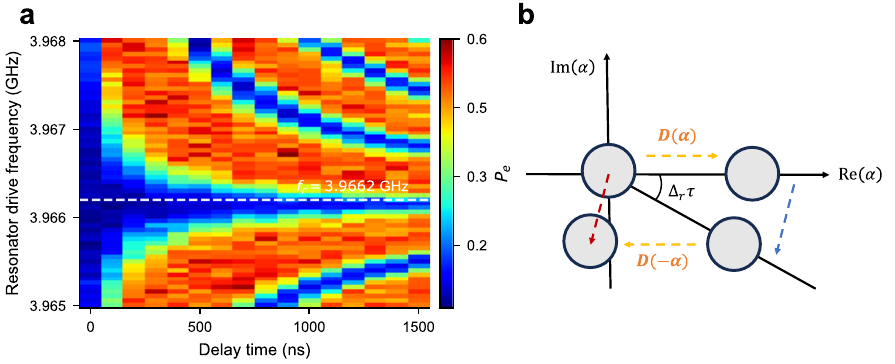}
        \caption{\textbf{Calibration of the resonator frequency.} \textbf{a}, Phase-to-displacement calibration data measured as a function of delay time and resonator drive frequency. The residual resonator population is minimized when the drive frequency matches the resonator frequency, and the dashed line marks the calibrated value $f_r=\omega_r/2\pi=3.9662$~GHz. \textbf{b}, Schematic of the calibration protocol: a coherent displacement is prepared, allowed to accumulate detuning-dependent phase, and then displaced back toward the vacuum.}
        \label{fig:resonator_frequency_calibration}
    \end{figure}

    The resonator frequency is calibrated using a phase-to-displacement sequence. Starting from the vacuum, we apply a calibrated displacement $D(\alpha)$, let the resonator evolve for a delay time $\tau$ in the frame of a trial drive frequency $\omega_d$, and then apply the opposite displacement $D(-\alpha)$. For a detuning $\Delta_r=\omega_r-\omega_d$, the coherent state evolves as
    \begin{equation}
        \ket{\alpha}\rightarrow\ket{\alpha e^{-i\Delta_r\tau}},
        \qquad
        \ket{\psi_f}\simeq
        \ket{\alpha(e^{-i\Delta_r\tau}-1)} ,
    \end{equation}
    up to an overall phase. The residual photon number after the second displacement is therefore
    \begin{equation}
        \langle n_f\rangle
        =
        |\alpha|^2\left|e^{-i\Delta_r\tau}-1\right|^2
        \simeq
        |\alpha|^2(\Delta_r\tau)^2 ,
    \end{equation}
    for $|\Delta_r\tau|\ll1$. The residual population is minimized when the trial drive frequency matches the resonator frequency. Experimentally, this residual resonator population is measured using the photon-number-resolved readout described above. The phase-to-displacement calibration data and schematic sequence are shown in Fig.~\ref{fig:resonator_frequency_calibration}.

    \subsubsection{Frame transformations}
    The synthesized AQRM is implemented from the tunable Jaynes-Cummings interaction between the system qubit and the resonator. The calibrated coupler pulse sets the bare exchange rate $g_0$, while the qubit control fields provide the longitudinal modulation, the carrier tone, and the transverse sideband tones used in the effective-Hamiltonian construction. In the rotating frame of the carrier, these controls determine the effective resonator frequency $\omega$, qubit frequency $\Omega$, rotating and counterrotating coupling strengths $g_1$ and $g_2$, and bias $\varepsilon$ defined in Eq.~\eqref{eqAQRM}.
    
    The first frame transformation introduced in the derivation corresponds to the qubit-resonator rotating frame. It is not implemented as an additional physical gate. Instead, the accumulated phase is tracked in software and absorbed into the phase references of subsequent qubit XY pulses, resonator displacements, and tomography rotations. The second transformation,
    \begin{equation}
        \hat{U}_2(t)=\exp\left[i\beta\sin(\nu t)\hat{\sigma}_x\right],
    \end{equation}
    is implemented as a parameterized rotation around the qubit $x$ axis. With the convention
    \begin{equation}
        X_{\theta}=\exp\left(-i\frac{\theta}{2}\hat{\sigma}_x\right),
    \end{equation}
    we write
    \begin{equation}
        \hat{U}_2(t)=X_{\theta_2(t)},
        \qquad
        \theta_2(t)=-2\beta\sin(\nu t),
    \end{equation}
    up to the sign convention of the microwave drive phase.
    
    Experimentally, the required single-qubit rotations are synthesized from calibrated $X_{\pi/2}$ pulses and virtual $Z$ gates~\cite{McKay2017}. The decomposition
    \begin{equation}
        X_{\theta}
        =
        Z_{-\pi/2}X_{\pi/2}Z_{\pi-\theta}X_{\pi/2}Z_{-\pi/2}
    \end{equation}
    uses only two finite-duration microwave pulses; the $Z$ rotations are implemented virtually by phase updates. More generally, a rotation in the equatorial plane can be written as
    \begin{equation}
        R_{\phi}(\theta)
        =
        \exp\left[-i\frac{\theta}{2}
        \left(\cos\phi\,\hat{\sigma}_x+\sin\phi\,\hat{\sigma}_y\right)\right]
        =
        Z_{\phi}X_{\theta}Z_{-\phi}.
    \end{equation}
    In particular, $R_y(\theta)=R_{\pi/2}(\theta)$, which is the parameterized single-qubit rotation used for the tunneling-state initialization below.
    
\begin{figure}[h]
        \centering
        \includegraphics[width=0.9\textwidth]{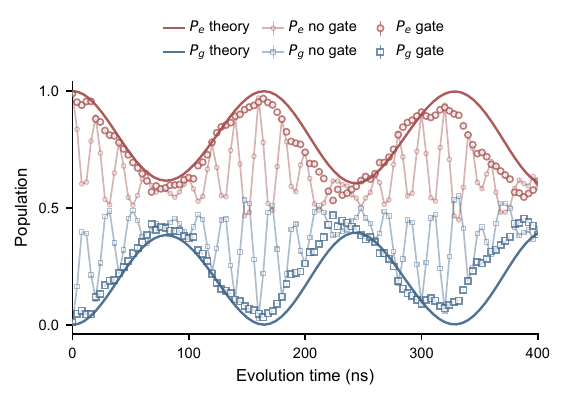}
        \caption{\textbf{Dynamical effect of the physical frame transformations.} Measured qubit-state populations during anisotropic Rabi dynamics with and without the two physical single-qubit operations used to implement the transformed-frame evolution. The gate data are shown as open markers with error bars, while the no-gate control data are shown as lighter marker lines to emphasize the faster oscillatory component. Solid curves show the numerical prediction for the transformed-frame dynamics. Error bars denote the standard error of the mean over ten repeated measurements.}
        \label{fig:dynamics_frame_transformation}
    \end{figure}

    Figure~\ref{fig:dynamics_frame_transformation} shows a dynamical check of the physical frame transformations. Applying the two single-qubit operations before and after the transformed-frame evolution changes the observed population dynamics in agreement with the numerical prediction, whereas omitting them gives a distinct control trajectory.
    
	\subsubsection{Photon-number-resolved resonator readout}
\begin{figure}[h]
	    \centering
	    \includegraphics[width=0.85\textwidth]{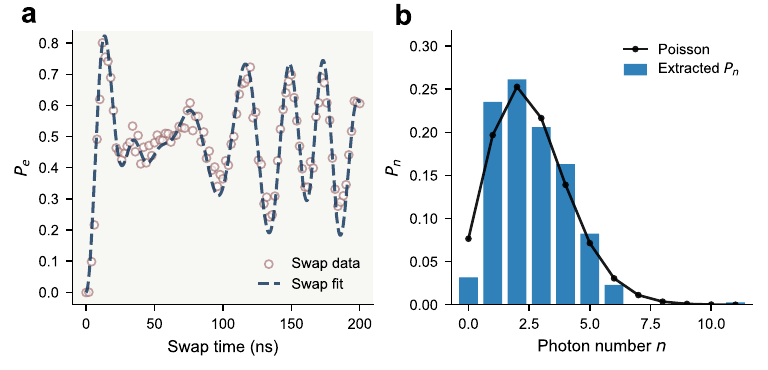}
	    \caption{\textbf{Photon-number-resolved resonator readout.} \textbf{a}, Representative resonant ancilla-resonator oscillation used to extract the photon-number distribution of the resonator state. The measured ancilla excitation probability is fitted with Eq.~\eqref{eq:photon_number_readout}. \textbf{b}, Extracted photon-number probabilities compared with a Poisson distribution with the same mean photon number, $\langle n\rangle=2.57$.}
	    \label{fig:photon_number_readout}
	\end{figure}

	The resonator photon-number distribution is measured with an ancillary qubit using resonant qubit-resonator swaps~\cite{LinPeng2013}. After the target evolution or calibration pulse, the resonator drive is switched off and the ancilla qubit $Q_a$, initially prepared in $\ket{g}$, is tuned into resonance with the resonator for a variable interaction time $\tau$. The ancilla is then returned to its idle frequency for projective readout. The resulting photon-number-dependent vacuum-Rabi oscillation is fitted to
	\begin{equation}
        P_e(\tau)=\frac{1}{2}\left[1-P_g(0)\sum_{n=0}^{n_{\mathrm{max}}}P_n e^{-\kappa_n\tau}
        \cos\left(2\sqrt{n}\,g_a\tau\right)\right],
        \label{eq:photon_number_readout}
    \end{equation}
    where $P_g(0)$ is the initial ground-state probability of the ancilla, $P_n=\rho_{nn}$ is the resonator photon-number distribution, $n_{\mathrm{max}}$ is the fitting cutoff, $\kappa_n$ is an empirical decay rate for the $n$-photon oscillation, and $g_a$ is the resonant ancilla-resonator coupling. In the experiment, $g_a/2\pi=11$~MHz. The extracted distribution gives the mean photon number $\langle n\rangle=\sum_n nP_n$ and is also used for the resonator-drive calibration and Wigner tomography below. A representative swap trace and the extracted photon-number distribution are shown in Fig.~\ref{fig:photon_number_readout}.
    
    \subsection{Preparation of ground state}
	
	\subsubsection{Adiabatic preparation of eigenstates}
\begin{figure}[h]
	    \centering
	    \includegraphics[width=0.95\textwidth]{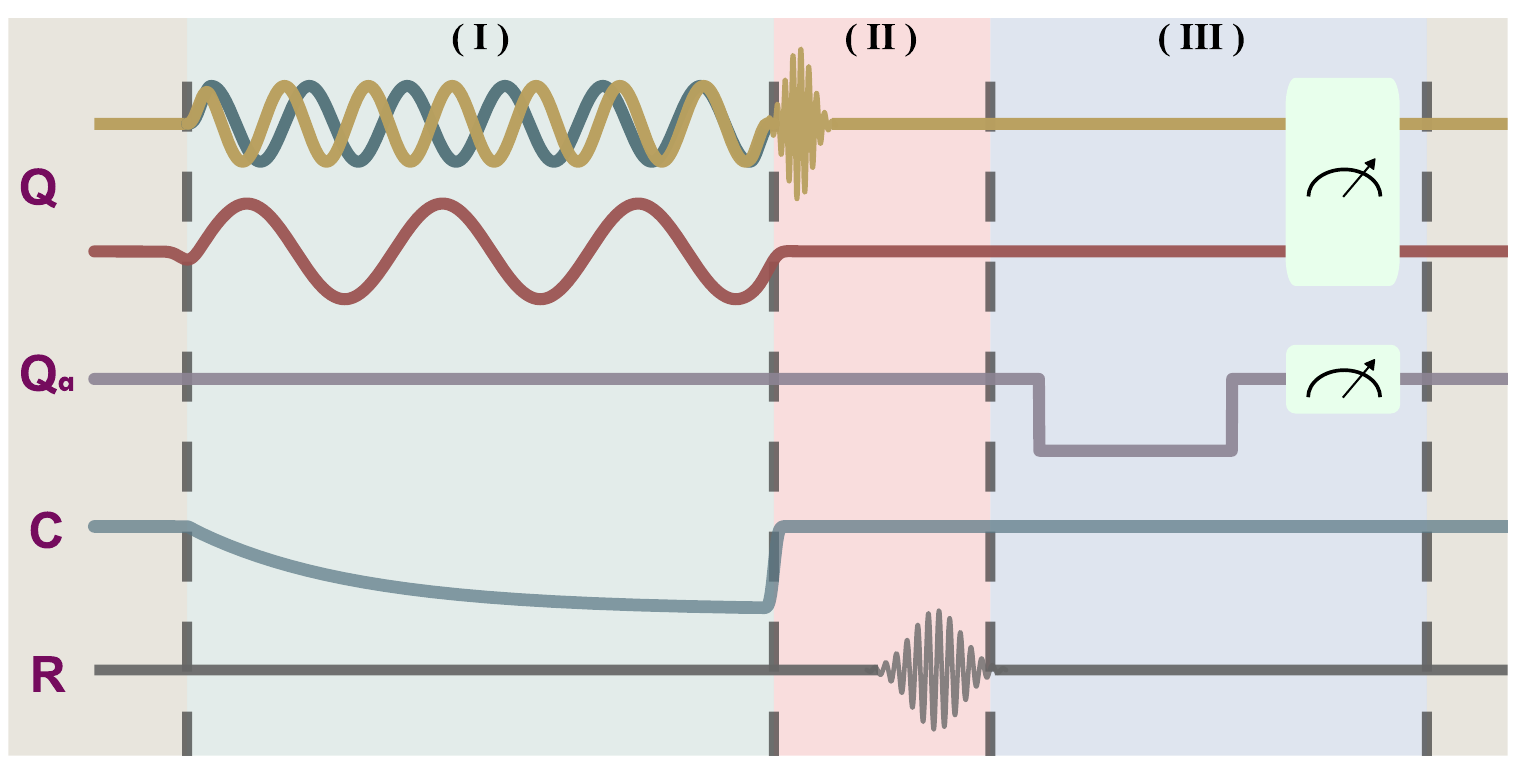}
	    \caption{\textbf{Adiabatic preparation of low-lying eigenstates.} \textbf{a}, Preparation and calibrated control pulses used to adiabatically connect decoupled product states to eigenstates of the target interacting Hamiltonian. The coupler-Z waveform implements the calibrated ramp $g_0(t)$, while the qubit and resonator control pulses set the effective Hamiltonian parameters. \textbf{b}, Mapping operations, including tomography rotations, resonator displacements, and ancilla-assisted photon-number mapping. \textbf{c}, Final measurement used to record the mapped qubit state.}
	    \label{fig:adiabatic_eigenstate_preparation}
	\end{figure}

	To resolve the low-lying spectral branches discussed in the main text, we prepare approximate eigenstates of the target Hamiltonian by adiabatically connecting simple product states to the interacting regime. The pulse sequence is summarized in Fig.~\ref{fig:adiabatic_eigenstate_preparation}. The sequence is organized into three experimental blocks: preparation and calibrated control pulses, photon-number and tomography mapping operations, and final measurement. In the preparation block, the resonator is initialized close to $\ket{0}$ and the system qubit is prepared in either $\ket{g}$ or $\ket{e}$, corresponding to the two decoupled product states used to access the two lowest branches. The calibrated coupler-Z waveform then ramps the effective qubit-resonator coupling to the target value while the qubit and resonator control fields set the synthesized Hamiltonian parameters.
	
	The ramp is described by
	\begin{equation}
	    g_0(t)=g_0^{\rm tar} f(t/T_{\rm ramp}),
	    \qquad
	    f(0)=0,\quad f(1)=1,
	    \label{eq:adiabatic_ramp}
	\end{equation}
	where $g_0^{\rm tar}$ is the target interaction strength and $f$ is a smooth ramp envelope obtained from the inverse tunable-coupling calibration. In the experiment, $T_{\rm ramp}$ is chosen for each target $g_0^{\rm tar}$ by balancing nonadiabatic leakage and decoherence, with typical values in the range of $1$--$3~\mu$s. After the ramp, the mapping block implements the joint readout basis: qubit rotations and resonator displacement pulses set the tomography basis, and the auxiliary-qubit--resonator interaction maps the photon-number populations onto the auxiliary qubit. The final measurement block records the corresponding qubit state by single-shot readout. The energy of the prepared branch is extracted from the reconstructed density matrix as
	\begin{equation}
	    E_{\rm exp}=\mathrm{Tr}\left(\rho_{\rm exp}H_{\rm tar}\right),
	    \label{eq:adiabatic_energy}
	\end{equation}
	where $\rho_{\rm exp}$ is the reconstructed density matrix and $H_{\rm tar}$ is the target Hamiltonian at the final parameters.
	
	The choice of ramp duration is guided by numerical simulations of the time-dependent effective Hamiltonian. For a target instantaneous eigenstate $\ket{\psi_j(g_0^{\rm tar})}$, we estimate the adiabatic-following fidelity
	\begin{equation}
	    F_j^{\rm ad}(T_{\rm ramp})=
	    \left|\bra{\psi_j(g_0^{\rm tar})}\psi_j(T_{\rm ramp})\rangle\right|^2,
	    \label{eq:adiabatic_following_fidelity}
	\end{equation}
	where $\ket{\psi_j(T_{\rm ramp})}$ is the state obtained by evolving the corresponding decoupled product state under the ramped Hamiltonian. The ramp durations used in the measurements are chosen in the plateau region of this numerical adiabatic-following estimate while remaining short compared with the relevant coherence times.
	
	\subsubsection{Joint qubit-resonator Wigner tomography}
    We reconstruct the joint qubit-resonator density matrix using displacement-assisted tomography~\cite{LinPeng2013}. Before the photon-number-resolved readout, we apply a single-qubit rotation $U_i\in\{I,R_x(\pi/2),R_y(\pi/2)\}$ to the system qubit and a displacement operation $D(\alpha_j)$ to the resonator. The resonator displacement maps information about off-diagonal resonator matrix elements onto measurable photon-number populations, while the qubit rotations select different qubit-basis components. By choosing a set of displacement points $\{\alpha_j\}$ and measuring the corresponding qubit-conditioned photon-number populations, the original joint density matrix is reconstructed from an over-constrained set of equations via least-squares optimization.
    
    Collecting the measured diagonal populations for all qubit rotations and resonator displacement points gives an over-constrained linear system,
    \begin{equation}
        \mathbf{y}_{\rm exp}=\mathcal{T}\mathbf{x},
        \qquad
        \mathbf{x}_{\rm LS}
        =
        \arg\min_{\mathbf{x}}
        \left\|\mathcal{T}\mathbf{x}-\mathbf{y}_{\rm exp}\right\|_2^2,
        \label{eq:joint_tomo_least_squares}
    \end{equation}
    where $\mathbf{x}$ is the vectorized joint density matrix and $\mathbf{y}_{\rm exp}$ contains the measured diagonal populations. In our measurements, the displacement points are chosen within a disk of radius $|\alpha|\leq1$, with ten displacement amplitudes used in the reconstruction. The Wigner matrix shown in the main text is calculated from the reconstructed joint density matrix.

    \subsection{Tunneling dynamics associated with hidden symmetries}
    
    \subsubsection{Preparation of displaced Fock states}
    The tunneling dynamics discussed in the main text is initialized in a product state of a displaced resonator state and a rotated qubit state. For the anisotropic case, the target initial state is
    \begin{equation}
        \ket{\psi_{\rm init}^{(\lambda)}}=
        D(-\alpha_{\lambda})\ket{0}\otimes\ket{+_{\lambda}},
        \qquad
        \alpha_{\lambda}=\frac{\sqrt{\lambda}g_1}{\omega},
        \qquad
        \ket{+_{\lambda}}=\frac{\sqrt{\lambda}\ket{g}+\ket{e}}{\sqrt{1+\lambda}}.
    \end{equation}
    Starting from $\ket{0,g}$, the qubit component $\ket{+_{\lambda}}$ is prepared by a single-qubit rotation $R_y(\theta_{\lambda})$ with
    \begin{equation}
        \theta_{\lambda}=2\arctan\left(\frac{1}{\sqrt{\lambda}}\right),
    \end{equation}
    followed by a resonator displacement pulse implementing $D(-\alpha_{\lambda})$. The parameterized rotation $R_y(\theta_{\lambda})$ is implemented using the single-qubit gate decomposition described above.
    
\begin{figure}[h]
        \centering
        \includegraphics[width=0.95\textwidth]{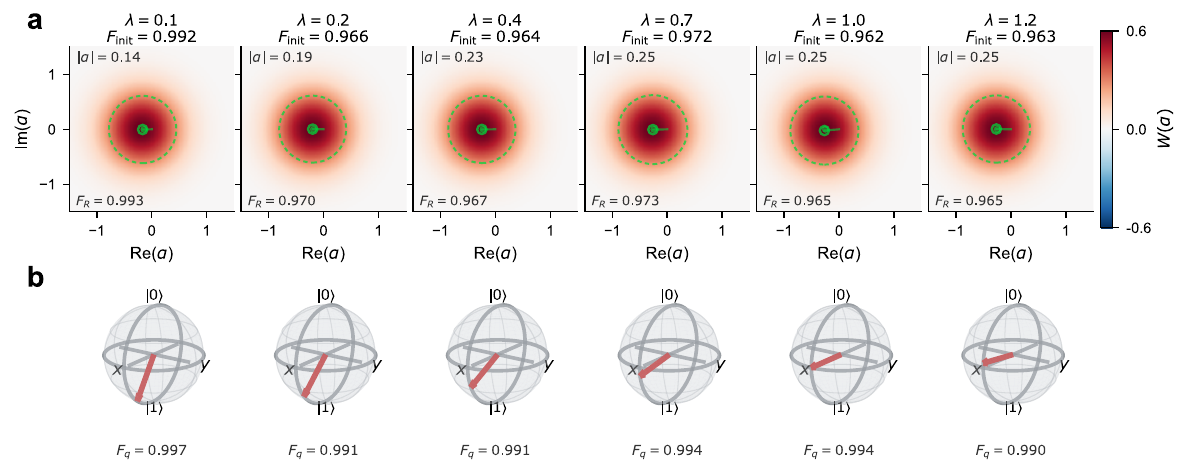}
        \caption{\textbf{Preparation and verification of tunneling-basis initial states.} \textbf{a}, Reduced resonator Wigner functions reconstructed from $\rho_R=\mathrm{Tr}_q(\rho_{\rm exp})$ for different anisotropy ratios $\lambda$. The resonator displacement is set by $|\alpha_\lambda|=\sqrt{\lambda}g_1/\omega$; the marked centers and contours indicate the reconstructed phase-space displacement. \textbf{b}, Qubit Bloch vectors reconstructed from the reduced qubit density matrices, with components $x=\mathrm{Tr}(\rho_q\hat{\sigma}_x)$, $y=\mathrm{Tr}(\rho_q\hat{\sigma}_y)$, and $z=\mathrm{Tr}(\rho_q\hat{\sigma}_z)$. The quoted fidelities are the joint initial-state fidelities $F_{\rm init}=\bra{\psi_{\rm init}^{(\lambda)}}\rho_{\rm exp}\ket{\psi_{\rm init}^{(\lambda)}}$, together with the reduced-state fidelities $F_R$ and $F_q$.}
        \label{fig:tunneling_state_preparation}
    \end{figure}

    The prepared state is verified by reconstructing the joint density matrix $\rho_{\rm exp}$ and evaluating
    \begin{equation}
        F_{\rm init}=
        \bra{\psi_{\rm init}^{(\lambda)}}\rho_{\rm exp}
        \ket{\psi_{\rm init}^{(\lambda)}}.
    \end{equation}
    For visualization, we trace out the qubit degree of freedom,
    \begin{equation}
        \rho_R=\mathrm{Tr}_q(\rho_{\rm exp}),
    \end{equation}
    and calculate the reduced resonator Wigner function from the reconstructed $\rho_R$. The reconstructed resonator Wigner functions and qubit Bloch vectors are shown in Fig.~\ref{fig:tunneling_state_preparation}.
    
    \subsubsection{Readout of displaced Fock states}
    The populations in the tunneling basis are measured by applying a state-mapping operation before the standard joint readout. For a target tunneling-basis state
    \begin{equation}
        \ket{\psi_{\rm tar}}=\mathcal{U}_{\rm map}\ket{n,m},
    \end{equation}
    where $\ket{n,m}$ denotes the product state of a bare resonator Fock state $\ket{n}$ and a qubit state $\ket{m}$ with $m\in\{g,e\}$, the population at time $t$ is
    \begin{equation}
        P_{\psi_{\rm tar}}(t)=
        \bra{n,m}
        \mathcal{U}_{\rm map}^{\dagger}\rho(t)\mathcal{U}_{\rm map}
        \ket{n,m}.
    \end{equation}
    Experimentally, this population is obtained by applying $\mathcal{U}_{\rm map}^{\dagger}$ before the joint readout. Here $\mathcal{U}_{\rm map}^{\dagger}$ consists of the inverse resonator displacement and the inverse qubit rotation. The population in the tunneling-basis state is then extracted from the photon-number-resolved resonator readout combined with the qubit-state readout.
    
	\clearpage
	\newpage
	
	\bibliography{sm}